\documentclass[a4paper,fleqn,usenatbib]{mnras}

\usepackage{newtxtext,newtxmath}
\usepackage[normalem]{ulem}

\usepackage[T1]{fontenc}
\usepackage{ae,aecompl}

\usepackage[pdftex]{graphicx}	
\usepackage{amsmath}	
\usepackage{xspace} 
\usepackage{booktabs} 

\newcommand*\blu[0]{\hbox{$\text{bl}_{\text{u}}$}\xspace}
\newcommand*\blo[0]{\hbox{$\text{bl}_{\text{o}}$}\xspace}

\newcommand*{\orcidlink}[1]{%
	\href{https://orcid.org/#1}{\includegraphics{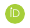}}%
}

\title[B/PS bulges and barlenses]{B/PS bulges and barlenses from a kinematic viewpoint. II}

\author[Daria Zakharova et al.]{Daria Zakharova$^{1,2}$\orcidlink{0009-0001-1809-4821}\thanks{E-mail: dzakharovaa@gmail.com},
Iliya S. Tikhonenko$^{3}$, Natalia Ya. Sotnikova$^{4}$, and Anton A. Smirnov$^{5}$\\
$^{1}$Dipartimento di Fisica e Astronomia "Galileo Galilei", Universita' degli studi di Padova, Vicolo dell'Osservatorio, 3, I-35122, Padova, Italy\\
$^{2}$INAF - Osservatorio astronomico di Padova, Vicolo dell'Osservatorio, 5, I-35122, Padova, Italy\\
$^{3}$Max-Planck-Institut für extraterrestrische Physik, Gießenbachstraße, D-85748 Garching, Germany\\
$^{4}$St. Petersburg State University,
Universitetskij pr.~28, 198504 St. Petersburg, Stary Peterhof, Russia\\
$^{5}$Central (Pulkovo) Astronomical Observatory of RAS, Pulkovskoye Chaussee 65/1, 196140 St. Petersburg, Russia\\
}

\date{Accepted XXX. Received YYY; in original form ZZZ}

\pubyear{2023}

\begin{document}
\label{firstpage}
\pagerange{\pageref{firstpage}--\pageref{lastpage}}
\maketitle

\begin{abstract}
    Internal dynamics and kinematics of galaxies have imprints on the line-of-sight velocity distribution~(LOSVD). Gauss-Hermite parametrisation allows one to identify the kinematics features of the system in terms of skewness~($h_3$) and broadness~($h_4$) deviations of a LOSVD. Such a method provides information about the type of orbits since a $h_3-\overline V$ correlation is a sign of elongated orbits, and the anti-correlation is a sign of circular or near-circular orbits. In previous works, analysis of the $h_3-\overline V$ relation provided a tool to identify a hidden bar or B/PS bulge~(edge-on, $\mathrm{PA}=90^\circ$) and to probe their strength. We prepared two $N$-body galaxy models with clear B/PS bulges: one has an ordinary bar~(the X model), and the second one has a barlens embedded into a bar~(the BL model) to investigate the mechanism of formation of $h_3$ features at any position of an observer. We show that the $h_3-\overline V$ correlation appears in the regions where bar and disc particles are mixing. We also reveal that the model with a barlens has an $h_3-\overline V$ anti-correlation in the centre, and we show that barlens-specific orbits are responsible for this signal. Moreover, this feature can be observed only for galaxies with compact bulges and barlenses. The results of this work are applicable for the interpretation of future Integral-field unit (IFU) data for real galaxies with B/PS bulges, especially for objects with barlenses.
\end{abstract}



\section{Introduction}
The diagnostics of galaxy kinematic features based on the \textcolor{black}{line of sight velocity distribution} (LOSVD) parameterization using the Gauss–Hermite series \citep{vanderMarel_Franx1993,Gerhard1993,Bureau_Athanassoula2005,Debattista_etal2005,Iannuzzi_Athanassoula2015,Li_etal2018} has become widespread, especially in IFU spectroscopic observations (for some IFU surveys and their results see, for example,  \citealp{Cappellari_etal2011,Sanchez_etal2012,Garcia-Benito_etal2015,vandeSande_etal2017,Bundy_etal2015,Gadotti_etal2019}). 
\textcolor{black}{The parameter $h_4$ of the LOSVD represents its symmetric deviation from a Gaussian profile~(broadness). $h_4$ reflects the features of the vertical density distribution for disc galaxies, visible almost face-on~(when only the $v_z$ velocity component is on the line of sight). This parameter is a good diagnostic of the B/PS bulge \citep{Debattista_etal2005}.}
The parameter $h_3$ (the skewness parameter) describes the asymmetric deviation from a Gaussian profile. Correlations or anti-correlations between $h_3$ and the mean line-of-sight (LOS) velocity ($\overline V$) of Gauss–Hermite series tell about the underlying stellar orbital structure and the rotation of different components \citep{Bureau_Athanassoula2005,Iannuzzi_Athanassoula2015,Li_etal2018}. 
\par
\citet{Bureau_Athanassoula2005} used a set of $N$-body models and showed that the 1D major-axis kinematics of edge-on galaxies is characterised by $h_3-\overline V$ anti-correlations in the disc region, while the bar exhibits positive correlations between these parameters with larger $h_3$ values in the end-on case than in the side-on case. \citet{Iannuzzi_Athanassoula2015} recovered and confirmed these results using a large set of barred models and 2D kinematic maps besides 1D cuts. They concluded that the presence of a strong positive $h_3-\overline V$ correlation seems to be related to the B/PS bulge (the thickest part of the bar), more than to the bar itself. Thus, this correlation helps, in addition to other methods, to study the imprints of the bar and boxy/peanut structures on the 2D line-of-sight kinematics of disc galaxies.
\par
The situation becomes more complicated at intermediate inclinations ($i=30-60^\circ$). In this case, other velocity components ($v_\varphi$ and $v_\mathrm{R}$, in addition to $v_z$ for the face-on case) are added to the LOSVD, so the diagnostics of the vertical structure in the bar area based on the parameter $h_4$ becomes ambiguous \citep{Zakharova_etal2023}. In addition, there is a redistribution of the contributions of the LOSVDs from various stellar components falling on the line of sight to the total LOSVD when we switch from the edge-on view to the view at intermediate inclinations \citep{Li_etal2018}. 
\par
In moderately inclined discs, $h_3-\overline V$ correlation in the bar area transforms into anti-correlation.
\citet{Iannuzzi_Athanassoula2015} found that these changes are induced very rapidly from $i=90^\circ$ to $i=75^\circ$ at the end-on view, and they do not practically evolve up to $i=60^\circ$. \citet{Li_etal2018} using two $N$-body models of disc galaxies with different bar strengths presented a more complicated picture for the inclination $i=60^\circ$ in which the outer parts of bars exhibit anti-correlations, while the core areas dominated by the B/PS bulges still maintain weak positive correlations. The anti-correlation in the outer bar area is explained by the complex superposition of LOSVDs of the bar and foreground/background disc.
The relationship between $h_3$ and $\overline V$ is crucial to correctly interpret the available IFU data.Therefore, it is necessary to deeply understand the reasons for such relationships for different regions of the galaxy and different inclinations.
\par
The first paper of series, \citet{Zakharova_etal2023} focused on the parameter $h_4$ for \textcolor{black}{several models} of galaxies with B/PS bulges at moderate inclinations and came to the conclusion that only for cuts along the major axis of the bar, the two-sided negative minima of this parameter carry information about the features (a `peanut') of the vertical density distribution, as in the face-on case (see \citealp{Debattista_etal2005}). These minima are associated either with x1 or with boxy orbits~(depending on the exact model). \textcolor{black}{\citet{Zakharova_etal2023} paid special attention to the model with  a barlens\footnote{Barlens is described as a new type of lens embedded into the bar and covering approximately half its length \citep{Laurikainen_etal2011}.}. There is an opinion that `barlens' and `B/PS' are the same stellar structure, namely the inner part of the bar, but seen at different viewing angles \citep{Laurikainen_etal2014}. In this case, the negative minima of $h_4$ should be associated specifically with a barlens. However, \citet{Zakharova_etal2023} showed that barlens} itself does not contribute to $h_4$ minima for the corresponding model, so we can not distinguish it using the information about vertical density. \textcolor{black}{Nevertheless}, the parameter $h_3$ may trace the central regions of the barlens.
\par
In this paper, we carry out a comparative analysis of the kinematic maps of the parameter $h_3$ for two models with different bar morphology, namely, for a galaxy with a barlens and with an ordinary bar. We compare maps for intermediate inclinations (mainly for $i=40^\circ$) and different viewing angles (hereinafter labelled PA) of the bar major axis relative to the line of nodes (LON). Based on the LOSVD in each pixel, we find characteristic patterns in the appearance of anti-correlations/correlations between $h_3$ and $\overline V$. We also show how the inner parts of the barlens show up on these maps and how to tell them apart from the inner discs.
\par
The paper is organised as follows.
In Section~\ref{sec:models}, we describe our $N$-body models, remind how we dissect them into various orbital groups in the bar, and how we prepare data cubes for all models for extracting the main kinematic parameters from LOSVDs in each pixel.
In Section~\ref{sec:h3_v}, we present our results comparing $h_3$ maps for models with barlens and with an ordinary bar at edge-on view and other positions of a bar. 
In Section~\ref{sec:barlens}, we focus on a central kinematic feature of our model with a barlens. 
In Section~\ref{sec:discussion}, we discuss the features of our $h_3$ maps for all models and astrophysical applications of our results. 
In Section~6, we give our conclusions.

\section{Data and models}
\label{sec:models}
\begin{figure}
   \centering
   \includegraphics[width=1\linewidth]{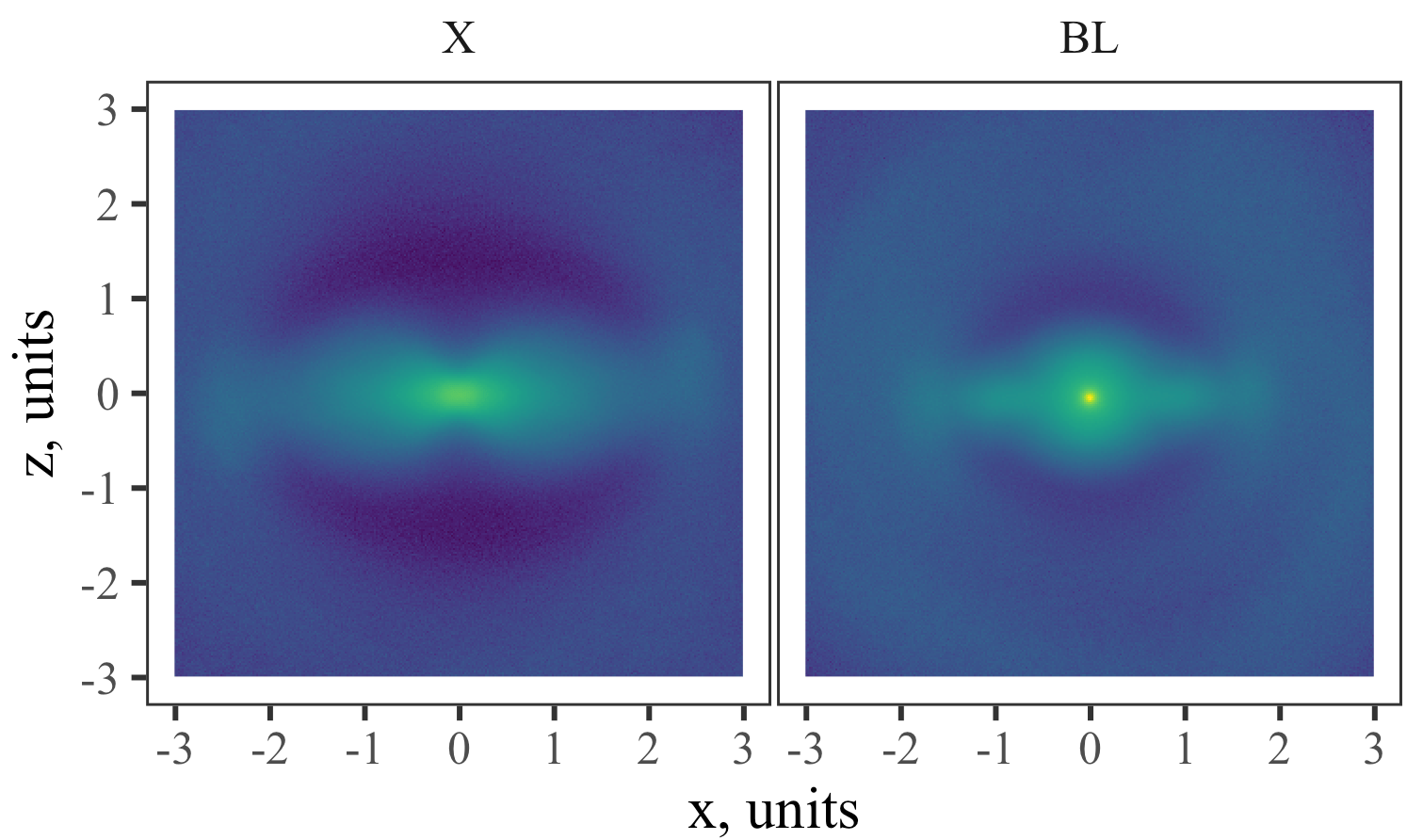}
   \caption{Face-on view of X and BL models in $(x,y)=[-3; 3]\times [-3;3]$ area.}
    \label{fig:faceon}%
\end{figure}

In this work, we employ the same grid of 4 self-consistent N-body models from \citet{Smirnov_etal2021}, which was used in the first part of the current study \citep{Zakharova_etal2023}. These models are initially composed of an exponential disc with an isothermal vertical profile, consisting of $4\times10^6$ particles, NFW-like ``live'' halo ($4.5\times10^6$ particles), and (except for one) a Hernquist bulge ($0.4\times10^6$ particles)  and then evolved using \texttt{gyrfalcON} integrator \citep{Dehnen2002} from the \texttt{NEMO} suite \citep{Teuben_1995}. 
For the purposes of this study, we mostly focus on two boundary cases: the model X (in terms of \citealt{Smirnov_etal2021}), which does not have any bulge component (\textcolor{black}{left} panel of Fig.\ref{fig:faceon}), and the BL model, hosting the most compact bulge of all with $M_\mathrm{b} = 0.1 M_\mathrm{d}$ and $r_\mathrm{b} = 0.05$ exponential scale lengths of the disk (\textcolor{black}{right} panel of Fig.~\ref{fig:faceon}). The exact parameters of the models can be found in \cite{Smirnov_etal2021}  or \cite{Zakharova_etal2023}.  \textcolor{black}{In Section~\ref{sec:discussion}, we also consider two other models (Xb and BLx) from \citet{Zakharova_etal2023} while analysing the $h_4$ maps.  Both the BL and the BLx models have a barlens morphology and a classical bulge with the same mass~($M_\mathrm{b}=0.1$), but the bulge of the BLx model is more rarefied (the bulge in the BL model has a radial scale $r_\mathrm{b} = 0.05$, while the bulge in the BLx model has $r_\mathrm{b} = 0.1$ ). At late stages of evolution, all of  the considered models develop clear B/PS bulges.}
\par
\citet{Smirnov_etal2021} applied frequency analysis techniques to the orbits in their models and identified the groups responsible for face-on bar morphology. They conduct their analysis in the bar frame and obtain three main Cartesian frequencies for each orbit ($f_x$, $f_y$, $f_z$) and a radial oscillation frequency $f_R$. Using this approach, they were able to separate the bar and the outer disk. The classification in terms of frequency ratios $f_x/f_R$ and $f_y/f_R$ allowed them to define several orbital groups constituting the bar: so-called `boxy bar', composed of boxy orbits with $f_R/f_x \approx 2$ and $f_y > f_x$, $x_1$-like (and $x_2$-like) group, characterized by $f_R/f_x \approx 2$ and $f_y \approx f_x$, and two non-classical orbital groups with $f_R/f_x>2$ and $f_R/f_x < 2$, labeled \blo and \blu, respectively.
In particular, they found that a key component of the barlens in the BL model is the \blu orbital group. While for the X model, most of its bar was found to be composed of boxy orbits, which explains its `face-on peanut' bar morphology.
\par
To decipher the kinematic features of the model, we re-use the datacubes made in the previous part of this study \citep{Zakharova_etal2023}. These datacubes are constructed by stacking several model snapshots near the midpoint of the simulation ($t=450$ time units), thus effectively reaching $\approx 7.5\times10^7$ particles in total and projecting them with different inclination angles ($i = 40^\circ, 60^\circ, 90^\circ$) and bar viewing angles ($\mathrm{PA} = 0^\circ, 45^\circ, 90^\circ$). The spatial resolution of the cubes matches the $300\times 300$ px MUSE grid. To calculate the kinematic parameters ($\overline{V}, \sigma, h_3, h_4$), we fit each LOSVD with Gauss-Hermite series up to the fourth order. The contributions of individual orbital groups in the total LOSVD are taken into account via a set of datacubes with orbital groups excluded one by one. We refer the reader to the first paper~\citep{Zakharova_etal2023} in this series for all additional details. 

\section{Correlations/anti-correlations between $h_3$ and $\overline V$}
\label{sec:h3_v}
In this section, we consider two models with clear and strong B/PS bulges but morphologically different face-on bars from two viewpoints: at edge-on position and an intermediate inclination~($i=40^\circ$) with the addition of other inclinations to understand
\begin{itemize}
\item[i)] how the relationship between $h_3$ and $\overline V$ appears in the bar region at inclinations less than $60^\circ$;
\item[ii)] how comparative analysis of two models makes it possible to distinguish two morphologically different bars on $h_3$ maps.
\end{itemize}

\subsection{Edge-on case}
\label{sec:edgeon}
\begin{figure*}
   \centering
   \includegraphics[width=1\linewidth]{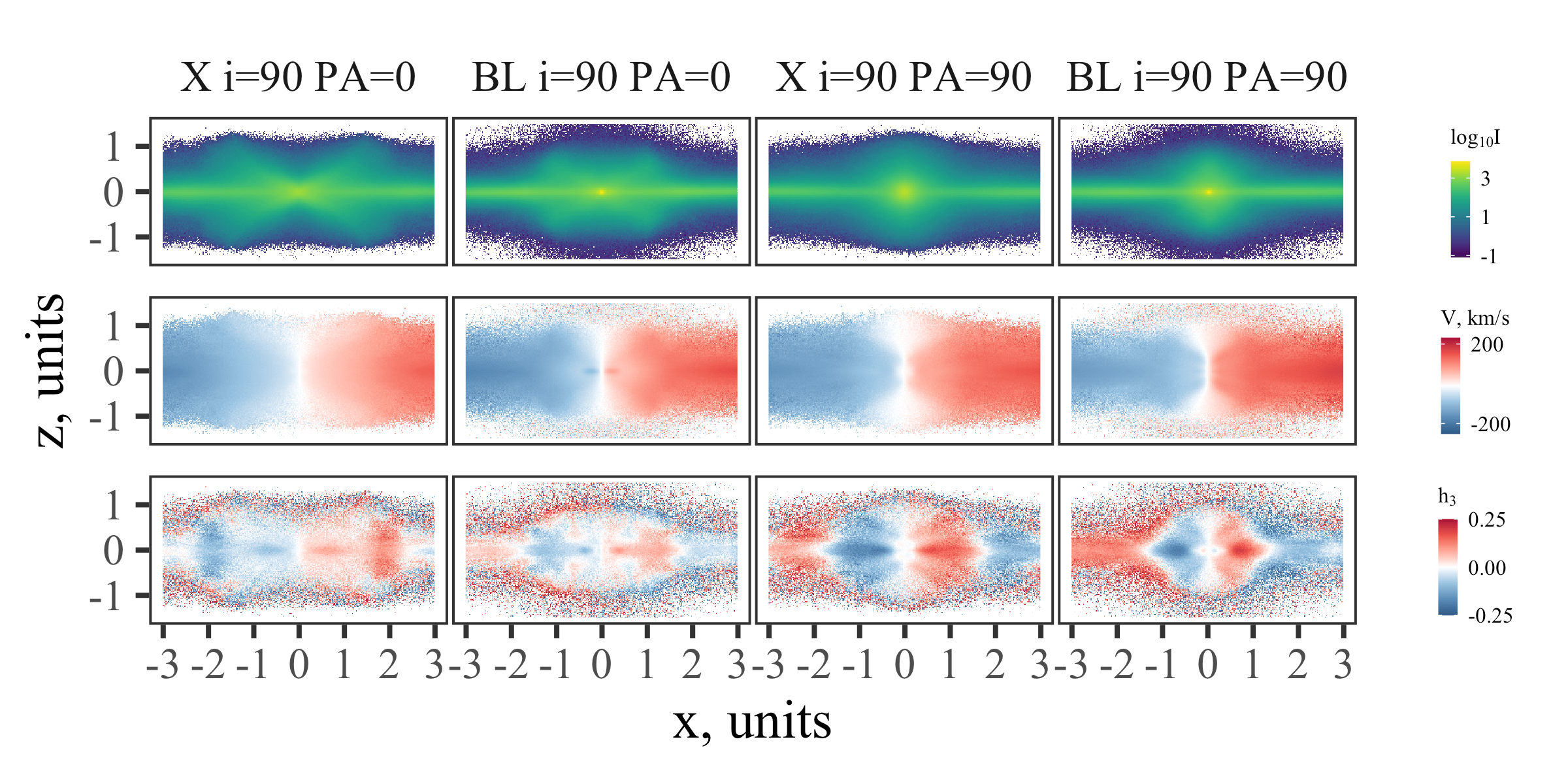}
   \caption{Intensity (top row), mean LOS velocity $\overline V$ (middle row) and $h_3$ (bottom) maps for X and BL models viewed edge-on ($i=90^\circ$) with different bar viewing angles ($\mathrm{PAs}$). The first two columns represent the bar with a major axis perpendicular to the line of sight~($\mathrm{PA}=0^\circ$, end-on view), and third-fourth columns show the bar with a major axis along the line of sight~($\mathrm{PA}=90^\circ$, side-on view). The disc rotates anticlockwise.}
    \label{fig:edge_on_maps}%
\end{figure*}

Fig.~\ref{fig:edge_on_maps} shows $h_3$ maps for two models with B/PS bulges at the edge-on view, the X model with an ordinary bar and the BL model with a barlens. Here we focus on two bar positions when the B/PS bulge and the X-structure associated with it are clearly seen~($\mathrm{PA}=0^\circ$, side-on view, the bar major axis is perpendicular to the line of sight, the first and second columns of Fig.~\ref{fig:edge_on_maps}) and also when the bar can not be identified~($\mathrm{PA}=90^\circ$, end-on view, the bar major axis goes along the line of sight, third and fourth columns of Fig.~\ref{fig:edge_on_maps}).
\par
In accordance with previous works~\citep{Bureau_Athanassoula2005, Iannuzzi_Athanassoula2015}, the disc area exhibits anti-correlations between $h_3$ and $\overline V$ while the bar area at any $\mathrm{PA}$ is characterised by positive $h_3-\overline V$ correlations. We note it for both our models~(Fig.~\ref{fig:edge_on_maps}). This positive correlation persists at least to the outermost boundaries of the B/PS bulge. Moreover, this correlation reflects the bar properties better than intensity maps in the case of a hidden bar~(for instance, when $\mathrm{PA}=90^\circ$, the $h_3$ maps show more differences between bars of the X and BL models than intensity maps).
\par
\begin{figure}
   \centering
   \includegraphics[width=1\linewidth]{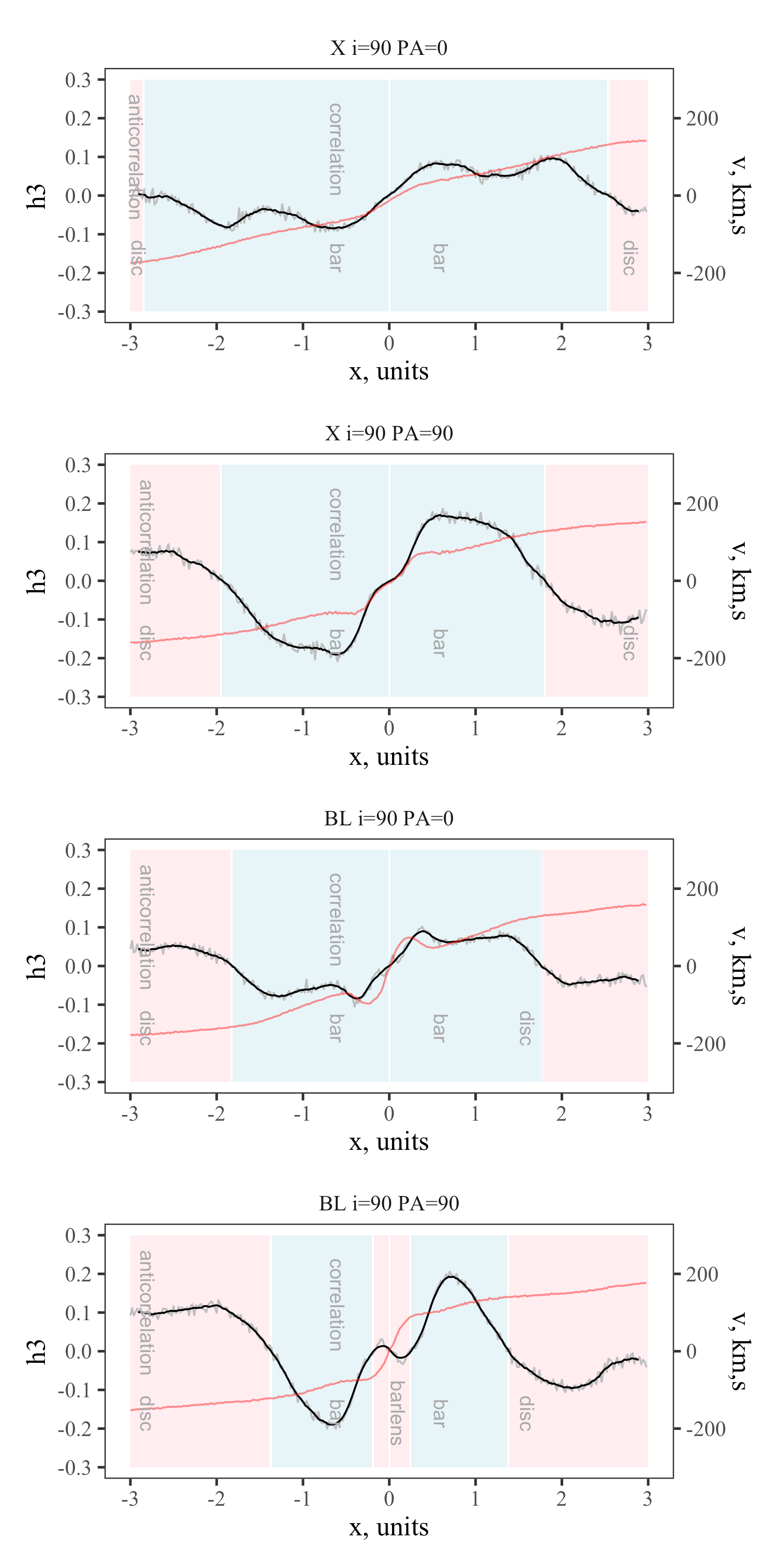}
   \caption{The $h_3$ and mean LOS velocity $\overline V$ (red line) profiles along $z=0$ for X and BL models at edge-on position with $\mathrm{PA}=0^\circ$ and $\mathrm{PA}=90^\circ$. The black line is rollmeaned $h_3$ values~(gray line shows exact values). The background colors represent areas of $h_3-\overline V$ correlation~(light blue) or anti-correlation~(red). Each zone is labelled with the name of a leading component of the (anti-)correlation.}
    \label{fig:edge_on_cuts}%
\end{figure}

We also note one feature of our BL model, which was not detected on the $h_3$ maps in previous works. Unlike the X model, the BL model demonstrates a clear anti-correlation between $h_3$ and $\overline V$ in the innermost regions at $\mathrm{PA}=90^\circ$. Central anti-correlation is also noticeable on the cuts of $h_3$ and $\overline V$ along $z=0$ ($h_3$ rises to the positive values on the left and falls to the negative values on the right, Fig.~\ref{fig:edge_on_cuts}). There is a hint of such a feature in the centre (two-sided small peaks\footnote{We note that rollmeaning of $h_3$ along the $z=0$ erases these peaks (Fig.~\ref{fig:edge_on_cuts}).})
for $\mathrm{PA}=0^\circ$, but the sign of $h_3$ remains unchanged. Moreover, on the $h_3$ maps for $\mathrm{PA}=0^\circ$, this feature is not visible (Fig.~\ref{fig:edge_on_maps}). At $\mathrm{PA}=90^\circ$, we attribute this feature (central anti-correlation) to specific orbits assembled into barlens at the centre (see Section~\ref{sec:barlens}).


\subsection{Inclined galaxies}
\label{sec:inclination}
\begin{figure*}
   \centering
   \includegraphics[width=1\linewidth]{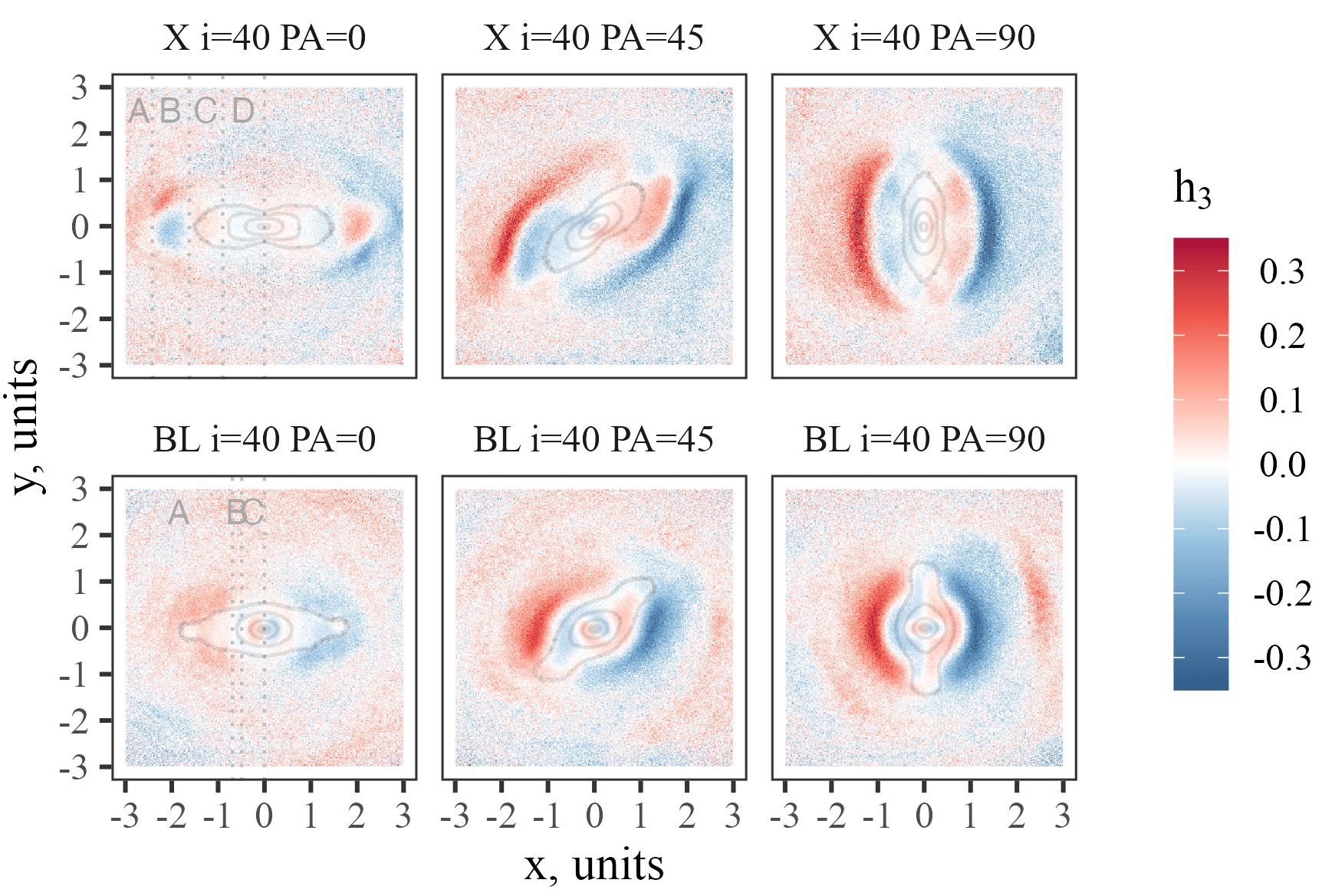}
   \caption{The $h_3$  maps for X~(top panel) and BL~(bottom panel) models at $i=40^\circ$ with different bar viewing angles ($\mathrm{PAs}$). Corresponding $\overline V$ maps are at the top right corner of each $h_3$ map. The LON coincides with the $x$ axis. The isophotes for each model correspond to the intensity maps at $i=40^\circ$. \textcolor{black}{The stripes A, B, C, and D highlight zones of correlation and anticorrelation. The lines at the boundary of the zones passes through $h_3\approx 0$ on the bar major axis. The similar notation is used in Fig.~\ref{fig:bl_x_h3_i=40_cuts} and Fig.~\ref{fig:x_velocityanalysis_for_discussion}.}}
    \label{fig:bl_x_h3_i=40}%
\end{figure*}

\begin{figure}
   \centering\includegraphics[width=1\linewidth]{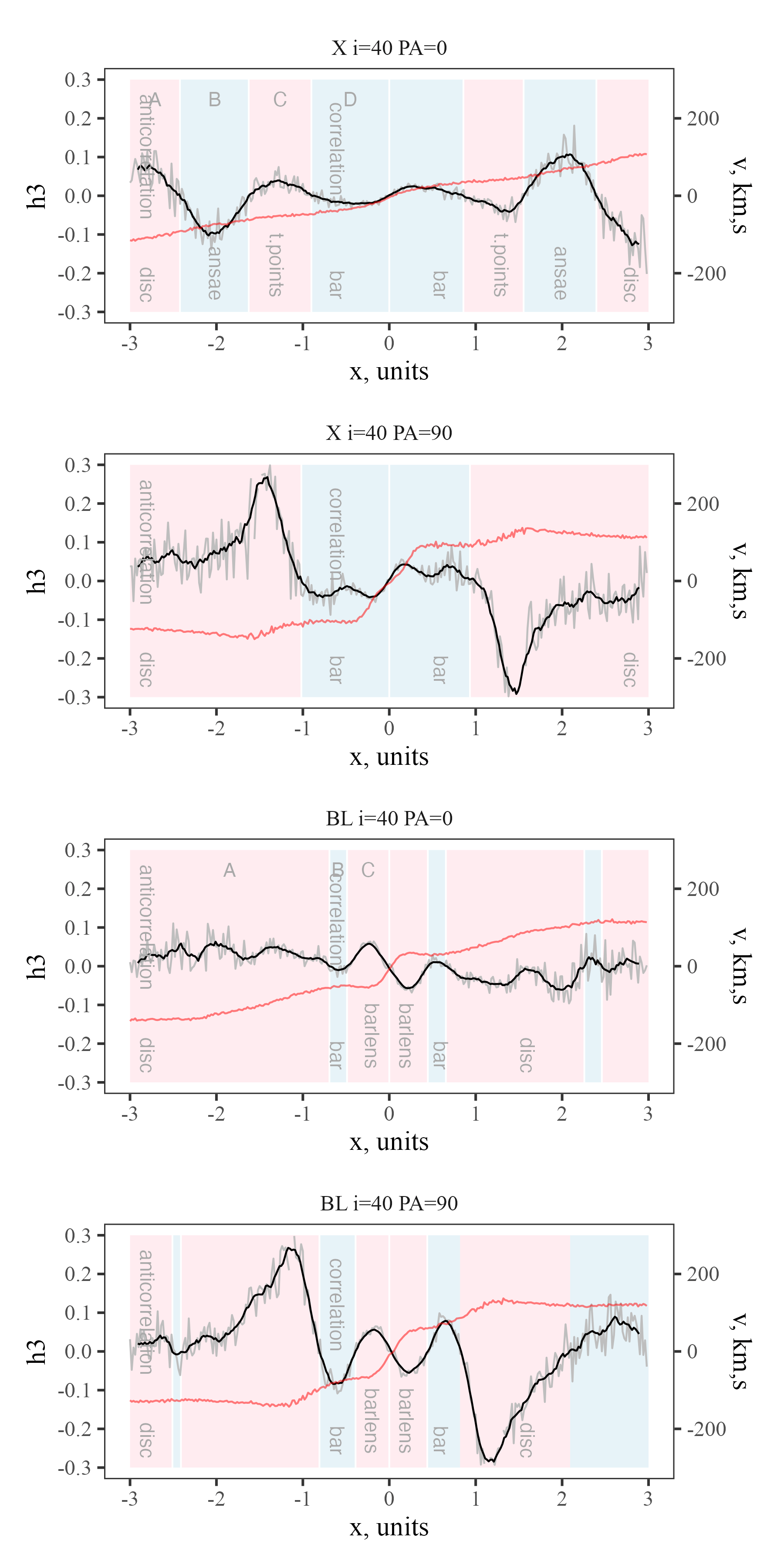}
   \caption{The same as in Fig.~\ref{fig:edge_on_cuts} but for $i=40^\circ$. \textcolor{black}{The vertical lines and the labels~(A, B, C or D) on the top panel convey the same meaning as in Fig.~\ref{fig:bl_x_h3_i=40}}.
   }
   \label{fig:bl_x_h3_i=40_cuts}%
\end{figure}
The $h_3-\overline V$ relations described above also appear at \textcolor{black}{intermediate} inclinations. Fig.~\ref{fig:bl_x_h3_i=40} presents $h_3$ maps for the inclination $i=40^\circ$ and different bar $\mathrm{PAs}$ for \textcolor{black}{our} two models~(plots \textcolor{black} for other inclinations and $\mathrm{PAs}$  can be found in the Appendix~\ref{app:h3maps_full_inclinations}). \textcolor{black}{At any inclination, $h_3$ maps reflects the sizes of the bar~(the X model has a bigger bar than the BL model)}.
\par
For the X model and $\mathrm{PA}=0^\circ$, 
at $i=40^\circ$, the outer parts  ($|x| \approx 1.5$) of bars exhibit anti-correlations~(\textcolor{black}{stripe C}), while the innermost areas still maintain correlations~(\textcolor{black}{stripe D}). Such a picture is shown both on the $h_3$ maps (Fig.~\ref{fig:bl_x_h3_i=40}) and on the cuts along the major axis (Fig.~\ref{fig:bl_x_h3_i=40_cuts}). Thus, for our X model, we note bar~(\textcolor{black}{stripe C in Fig.~\ref{fig:bl_x_h3_i=40} and Fig.~\ref{fig:bl_x_h3_i=40_cuts}}) and disc~(\textcolor{black}{stripe A in Fig.~\ref{fig:bl_x_h3_i=40} and Fig.~\ref{fig:bl_x_h3_i=40_cuts}}) signatures on $h_3$ maps for galaxies with intermediate inclinations (see also Fig.~\ref{fig:app_x_h3_full}).
Our X model is similar to one of the \citet{Li_etal2018} models (it has no bulge), so we observe a similar change from correlations to anti-correlations  in the outer parts of the B/PS bulge as the inclination decreases. 
\par
However, our X model shows an additional feature that was not visible on the $h_3$ maps/cuts in previous works. For all $\mathrm{PAs}$, we see ``spots'' with positive correlations between $h_3$ and $\overline V$ in the outermost areas of the bar (at its ends) along the line of nodes (LON) as in the edge-on case~(\textcolor{black}{stripe B in Fig.~\ref{fig:bl_x_h3_i=40} and Fig.~\ref{fig:bl_x_h3_i=40_cuts}}). We attribute the feature of these areas to the mixing of the disc and B/PS bulge particles. We discuss it in detail in Section~\ref{sec:x_model_discussion}.
\par
At $\mathrm{PA}=90^\circ$ in the area of the bar (B/PS bulge)~(\textcolor{black}{bar zone Fig.~\ref{fig:bl_x_h3_i=40_cuts}}) along the LON, we note a positive correlation, which in the transition region from the B/PS bulge to the disc, at the very end of the B/PS, is replaced by anti-correlation (\textcolor{black}{disc zone in } Figs.~\ref{fig:bl_x_h3_i=40}-\ref{fig:bl_x_h3_i=40_cuts}), as in the models by \citet{Li_etal2018}. The same behavior is observed for $i=20^\circ$ and $i=60^\circ$ (Fig.~\ref{fig:app_x_h3_full}).
\par
The change of correlations to anti-correlations in the area of the bar when moving from the edge-on case to the inclined one was attributed by \citet{Li_etal2018} to the weakening of the influence on the LOSVD of the foreground/background disc particles that fall into the line of sight at moderate inclinations.
\par
For the BL model, the common pattern of $h_3$ maps at $\mathrm{PA}=0^\circ$ and $\mathrm{PA}=90^\circ$ for all considered $i$ is the thin ring covering the outer regions of the barlens with a correlation between $h_3$ and $\overline V$~(\textcolor{black}{stripe B in Fig.~\ref{fig:bl_x_h3_i=40} and Fig.~\ref{fig:bl_x_h3_i=40_cuts} for the BL model} ). At $\mathrm{PA}=90^\circ$ one can also see an area at the ends of the B/PS bulge (barlens) where the anti-correlation between $h_3$ and $\overline V$ appears again (Figs.~\ref{fig:bl_x_h3_i=40}-\ref{fig:bl_x_h3_i=40_cuts} for $i=40^\circ$ and Fig.~\ref{fig:app_x_h3_full} for other inclinations). At $\mathrm{PA}=0^\circ$ the bar itself also exhibits~(\textcolor{black}{stripe B in Fig.~\ref{fig:bl_x_h3_i=40} and Fig.~\ref{fig:bl_x_h3_i=40_cuts} for the BL model} ) anti-correlation as in the model X. But for $i=60^\circ$ and $\mathrm{PA}=0^\circ$ (Fig.~\ref{fig:app_x_h3_full}), the $h_3$ map demonstrates additional spots of correlation at the very end of the bar as in the X model.
\par
For the BL model, $h_3$ maps have a central feature that radically distinguishes them from the maps for the X model for all considered $i$ and $\mathrm{PA}$~(\textcolor{black}{stripe C in Fig.~\ref{fig:bl_x_h3_i=40} and Fig.~\ref{fig:bl_x_h3_i=40_cuts} for the  BL model} ). The central area, where part of the barlens is localised, demonstrates the anti-correlation between $h_3$ and $\overline V$ as if we were observing a rotating disc in this region. This feature is especially well seen on the cuts along the LON (Fig.~\ref{fig:bl_x_h3_i=40_cuts}). Moreover, even in the most perturbed case ($\mathrm{PA}=45^\circ$), the innermost area is distinguished by $h_3-\overline V$ anti-correlation. We discuss this feature in the next section in detail.
\par
The $h_3$ maps for both our models are perturbed when the galaxy is rotated relative to the observer. The most distorted case is $\mathrm{PA}\approx 45^\circ$. However, in this case, the pattern of alternation of anti-correlation and correlation regions is similar to that observed at $\mathrm{PA}=0^\circ$ and $\mathrm{PA}=90^\circ$.
\par
Fig.~\ref{fig:bl_x_h3_i=40_cuts} summarizes the kinematic features associated with the parameter $h_3$ and the mean LOS velocity $\overline V$ for our two models, showing them in the cuts along the LON. The background colours in the figures highlight the $h_3-\overline V$ correlation/anti-correlation zones by blue/red colours along the LON. The figure demonstrates the effect of viewing angle on the bar signature $h_3-\overline V$ correlation in the centre for both models with an ordinary bar~(X) and the model with a barlens embedded in a bar~(BL). We note that the bar signature on $h_3$ maps depends not only on the bar size but on the inclination and bar viewing angle.

\section{Central features of the barlens model}
\label{sec:barlens}
\begin{figure}
   \centering
   \includegraphics[width=1\linewidth]{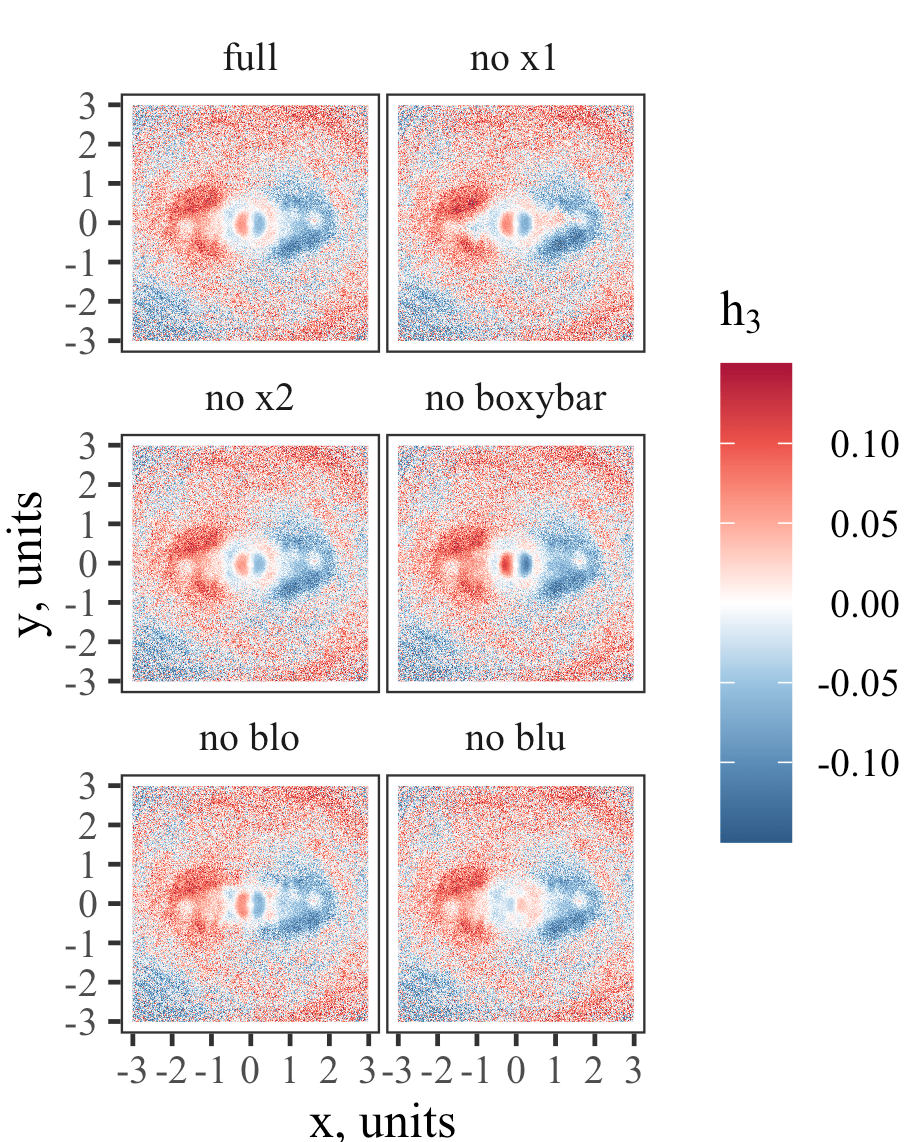}
   \caption{The $h_3$ maps for the BL model at $i=40^\circ$ and $\mathrm{PA}=0^\circ$ with excluded different bar orbital families~(the name of the excluded family is shown on the top of each panel). }
   \label{fig:bl_i40pa0_deleteorbits}%
\end{figure}

\begin{figure*}
   \centering
   \includegraphics[width=1\linewidth]{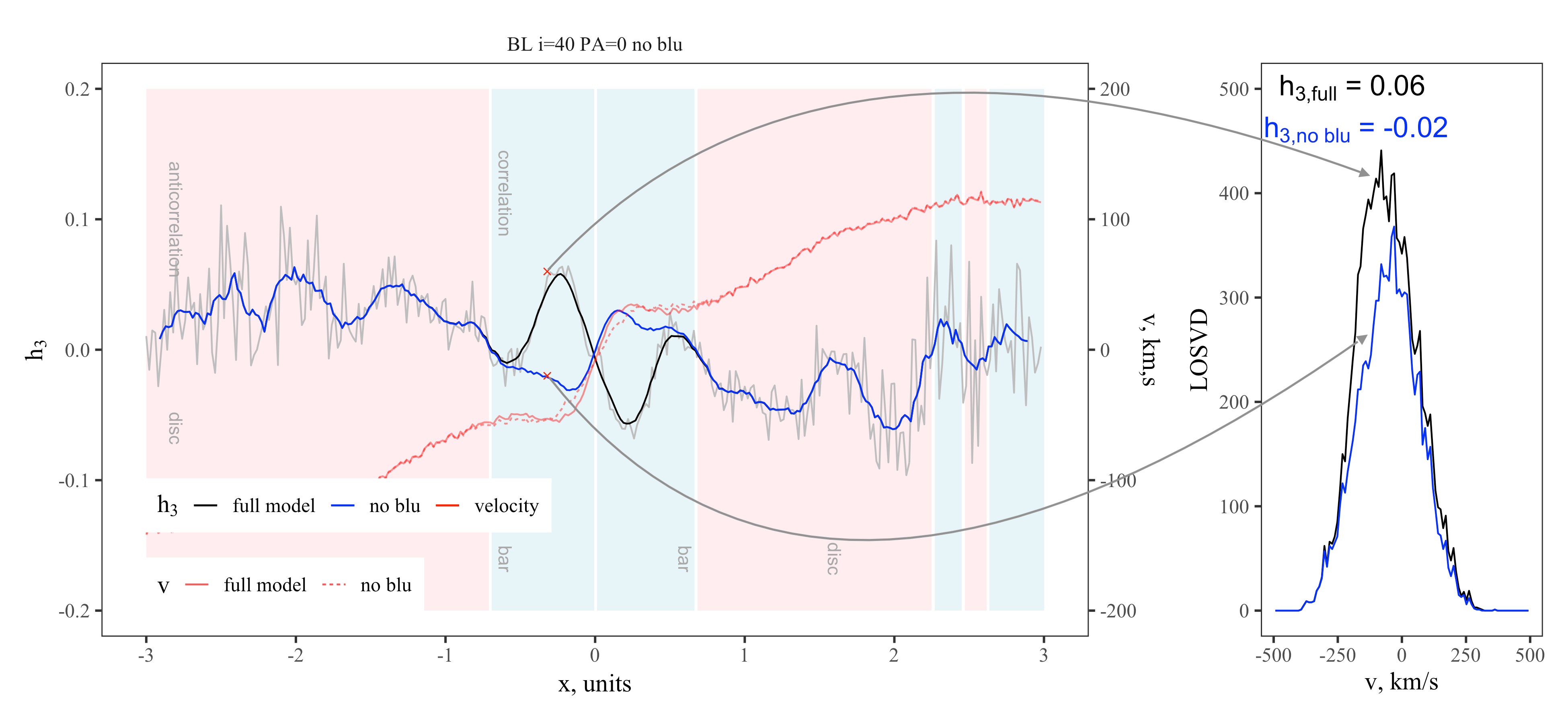}
   \caption{\textit{Left:} $h_3$ cuts along the LON for the full BL model~(gray line goes along the exact values and black line corresponds to the rollmeaned values) and for the BL model without \blu orbits~(blue line). The red line is $\overline V$ profile. The red~(light blue) zones highlight $h_3-\overline V$ anti-correlation (correlation) for the BL model without \blu orbits. Each zone is labelled with a name of component that mainly contributes to the $h_3$ value. \textit{Right:} LOSVDs in the same pixel for the full and no-\blu BL model~(the position of the pixel is shown on the left panel).}
    \label{fig:bl_i40pa0_deleteorbits_nod}%
\end{figure*}
In this section, we focus only on the BL model to understand how a barlens is kinematically imprinted on $h_3$ maps and how barlens creates \textcolor{black}{a distinctive  feature of this model}
~($h_3-\overline V$ anti-correlation in the central region). This feature distinguishes this model from all others. The traces of this anti-correlation can be found even on edge-on maps (Fig.~\ref{fig:edge_on_maps}). 
\par
This anti-correlation exists regardless of the viewing angle~(see Fig.\ref{fig:app_bl_h3_full}). It can be attributed mainly to any central substructure because, in general, in the innermost parts of the model, at moderate inclinations, mainly central particles fall into the line of sight. Correspondingly, this area is not much contaminated by the disc or even B/PS bulge particles. 
\par
The question arises whether the central $h_3-\overline V$ anti-correlation is related to the barlens. To establish it, we sequentially remove the main orbital groups from the model one at a time, rebuild our cubes, and recalculate the $h_3$ maps. The description of the orbital groups can be found in Section~\ref{sec:models}. Fig.~\ref{fig:bl_i40pa0_deleteorbits} shows the result of these procedures for the BL model, $i=40^\circ$ and $\mathrm{PA}=0^\circ$. According to this figure, the main contribution to the central region is due to \blu orbits~(rosette-like orbits constituting the barlens, \citealp{Smirnov_etal2021}). Fig.~\ref{fig:bl_i40pa0_deleteorbits_nod} demonstrates this point in more detail. The left panel of Fig.~\ref{fig:bl_i40pa0_deleteorbits_nod} shows the cuts along the LON for the total BL model~(black line, see also Fig.~\ref{fig:bl_x_h3_i=40_cuts})
and for the BL model with the \blu component excluded~(blue line). Central anti-correlation can be eliminated by removing \blu orbits, and then only the correlation characteristic of the bar is visible. Removing this orbital group removes this kinematic feature at all viewing angles, even when viewed edge-on (Fig.~\ref{fig:app_bl_no_blu_h3_full}).
The right panel of Fig.~\ref{fig:bl_i40pa0_deleteorbits_nod} shows the LOSVD in one pixel for the original and modified model. It can be seen  that the left wing of the LOSVD is created by foreground/background particles that are slower than the particles of the \blu orbital group. Due to this, an anti-correlation and a positive value of $h_3$ arise (the model rotates anticlockwise). When the \blu orbits are removed, the LOSVD becomes more symmetric, and the parameter $h_3$ decreases, taking on a negative value (resulting in a weak correlation). \textcolor{black}{Such a removal also changes the mean velocity $\overline V$ since the resulting distribution~(black line in the left panel of Fig.~\ref{fig:bl_i40pa0_deleteorbits_nod} ) is the sum of the distributions from each orbital group with different mean velocity. Therefore, the exclusion of fast (or slow) rotating components relative to the remainder (disc + bar) invariably leads to a changing $\overline V$. This is shown in Fig.~\ref{fig:bl_i40pa0_deleteorbits_nod} by solid and dotted red lines.}
\par
So, the central $h_3-\overline V$ anti-correlation in the BL model is created by the barlens specific orbital group. The plots for other inclinations and bar viewing angles are shown in the Appendix~\ref{app:h3maps_bl_noblu_inclinations} (Fig.~\ref{fig:app_bl_no_blu_h3_full}) to support our conclusion. Even at $i=90^\circ$ and $\mathrm{PA}=90^\circ$, the change from anti-correlation to correlation in the central regions occurs due to the removal of \blu orbits. For $i=90^\circ$ and $\mathrm{PA}=0^\circ$, no anti-correlation in the centre was observed (Fig.~\ref{fig:edge_on_maps}), but when the \blu orbits are removed, the correlation in the central regions becomes more pronounced and is visible as bright spots (Fig.~\ref{fig:app_bl_no_blu_h3_full}, bottom right plot, blue central spot on the left and red one on the right).
\par
Although in the previous works of the series~\citep{Smirnov_etal2021, Zakharova_etal2023} we found that the BL model is formed by several orbital families~(\blu, \blo, $x_1$, boxy-orbits, etc. ), we do not obtain any significant differences in $h_3$  maps removing other orbital families.

\section{Discussion and comparison with real data}
\label{sec:discussion}
Despite the difference in bar morphology and B/PS bulge sizes, $h_3$ maps for our models for edge-on view show generally the same behavior. In the area of the bar and the B/PS bulge, as its \textcolor{black}{thickest part}, there is a correlation between $h_3$ and $\overline V$. It is in agreement with the results of previous works. However, for intermediate inclinations, especially for $i<60^\circ$, the $h_3$ maps for our models begin to show different patterns, which, apparently, is due to differences between the X and the BL models. For example, we note the persistence of a weak correlation in the central areas of the bar with decreasing inclination, as in the models by \citet{Iannuzzi_Athanassoula2015} and \citet{Li_etal2018}. However, for our X model, without barlens and with a strong B/PS bulge, we see that the correlation is preserved at the very ends of the bar, where the bar is adjacent to the disc. This feature has not been noted in previous works. Moreover, for our BL model at almost all viewing angles, we find a very expressive kinematic feature in the centre, the anti-correlation associated with the central barlens. Although there were models similar to the BL model in \citet{Iannuzzi_Athanassoula2015}, this feature did not stand out on their maps. Below, we discuss these features in more detail, as well as possible applications of our results to the available IFU data.

\subsection{X model and related discussion}
\label{sec:x_model_discussion}
\begin{figure}
   \centering
   \includegraphics[width=\linewidth]{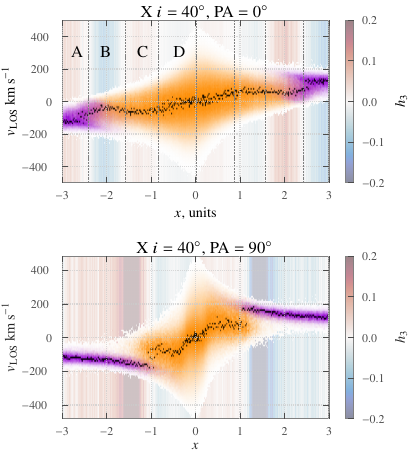}

   \caption{The position-velocity diagram (PVD) the X model viewed at $i=40^\circ$, $\mathrm{PA}=0^\circ$ (upper plot) and $\mathrm{PA}=90^\circ$ (bottom plot). The colored stripes on the background layer show the $h_3$ values along the LON, which corresponds to the major axis of the bar at $\mathrm{PA}=0^\circ$ and minor axis at $\mathrm{PA}=90^\circ$. In the foreground layer, in each column, the LOSVD in a single pixel of the slice is provided. The purple dots correspond to the LOSVDs of the disc particles, while the orange ones represent the LOSVDs of a bar. Each column is independently normalised by the maximum value of the total LOSVD in the corresponding pixel. In addition to that, the position of the maximum of each LOSVD is indicated by a black dot in each column. \textcolor{black}{The vertical lines and the labels on the top panel convey the same meaning as in Fig.~\ref{fig:bl_x_h3_i=40}}}
    \label{fig:x_velocityanalysis_for_discussion}%
\end{figure}

The alternating regions of correlations and anti-correlations along the LON at the bar ends for our X model at $i=40^\circ$ and $\mathrm{PA}=0^\circ$ (Fig.~\ref{fig:bl_x_h3_i=40}, top left panel) can be explained as following. If we start from the left side of the bar major axis (Fig.~\ref{fig:x_velocityanalysis_for_discussion}, upper panel), we first find a region of positive $h_3$ values in the disc region (the red stripe in Fig.~\ref{fig:x_velocityanalysis_for_discussion} \textcolor{black}{indicated with `A'}). These values correspond to the purple area (disc particles). The LOSVD for these particles is dominated by the fast-rotating disc particles at tangent points\footnote{Close to tangent points, the azimuthal velocity component or its projection onto the line of sight is the largest.} near the LON (the disc rotates anticlockwise), which gives a peak on the LOSVD on the side of negative velocity values (black dots in the figure). However, the LOSVD itself is slightly skewed towards positive values due to the minor contribution of foreground/background disc particles with a small projection of the rotation velocity on the line of sight. The foreground/background disc particles create a raised wing to the right of the LOSVD peak. It is expressed in a more extended distribution above the black dots in Fig.~\ref{fig:x_velocityanalysis_for_discussion} (above the distribution peaks) than below. As a result, we obtain anti-correlations \citep{Bureau_Athanassoula2005,Iannuzzi_Athanassoula2015,Li_etal2018}.
\par
Moving right along the LON, we encounter a region of negative $h_3$ values (\textcolor{black}{the B stripe} in Fig.~\ref{fig:x_velocityanalysis_for_discussion}). The mean value of $\overline V$ is also negative, but in this region, the situation is reversed. In this area, particles of the disc and bar are mixed (purple and orange colors are present at the same time). It is the region of the so-called ansae, and there are many bar particles in this feature. This region is mainly populated by orbits from x1v1 and x1v2 orbital families (following the notation by \citealp{Skokos_etal2002a}). They are narrow and not as elevated as boxy orbits in the region of the B/PS bulge (see, for example, figure~9 in \citealp{Parul_etal2020}). The LOSVD is dominated by bar particles at tangent points and areas adjacent to these points (black dots as LOSVD peaks are in the orange area). These particles rotate slower than the disc particles and create a peak at the LOSVD in the region of slowly rotating particles, i.e., to the right of the peak in the first case, in the region of anti-correlations. The rapidly rotating disc particles present here\footnote{In this region, disc particles are mainly present in the tangent points.} create a wing in the LOSVD on the left, on the side of negative velocity values that, in turn, results in correlations. The bright spots of correlations in the ansae region are more pronounced for $i=60^\circ$, but they are noticeable even at $i=20^\circ$ (Fig.~\ref{fig:app_x_h3_full}). Moreover, we found such spots on the $h_3$ maps for our BLx and Xb models that have a bar morphologically intermediate between the X and BL models (see \citealp{Smirnov_etal2021}). 
For the X model at $\mathrm{PA}=0^\circ$, in the region of ansae, \citet{Zakharova_etal2023} found a negative minimum of the parameter $h_4$ (their figure~A.3) for all considered inclinations. The negative values of $h_4$ are associated not with the features of the vertical density distribution but with the features of the LOSVD determined by the $v_\varphi$ velocity component. Here, the broadening of the LOSVD, leading to negative values of $h_4$, occurs due to the addition of the LOSVDs from two subsystems with different average rotation speeds (disc and bar).
\par
\textcolor{black}{In the C red stripe of Fig.~\ref{fig:x_velocityanalysis_for_discussion}, at $\mathrm{PA}=0^\circ$ particles of the bar} completely dominate (orange color), with practically no admixture of disc particles at such inclinations. The emergence of anti-correlations in this region for intermediate inclinations is explained as follows. In the moderately inclined cases, the lines of sight do not go through the outer disc areas, and the contribution of disc particles to the LOSVD can be neglected. Here, bar particles from tangent points and B/PS bulge particles from high altitudes fall on the line of sight. The main orbital family, which the 3D B/PS bulge is assembled from, is the boxy orbits \citep{Parul_etal2020,Smirnov_etal2021}. They create a wing on the LOSVD from the side of small absolute velocities. This wing skews the LOSVD formed by bar particles at tangent points and results in anti-correlations. For $\mathrm{PA}=90^\circ$, the reason for the burst of anti-correlation (parameter $h_3$ becomes abruptly positive, Fig.~\ref{fig:bl_x_h3_i=40_cuts}) when moving towards the centre is different. The disc and bar particles are mixed here (Fig.~\ref{fig:x_velocityanalysis_for_discussion}, bottom plot). The main peaks (black dots) on the LOSVD are associated with disc particles (purple area), and B/PS bulge particles from high altitudes give an extended low-velocity wing (orange). At $\mathrm{PA}=90^\circ$, this also works for $i=20^\circ$ and $i=60^\circ$ (Fig.~\ref{app:alternatingh3_LOSVDs_xbl}). The case of $i=60^\circ$ is particularly interesting because it can be used to identify the B/PS bulge when it is difficult to distinguish it on images since the line of sight goes through the entire X-structure. A sudden increase in anti-correlations followed by a sudden transition to a weak correlation means the end of the disc and the beginning of the bar, and the B/PS bulge in the area of correlation (see also \citealp{Li_etal2018}). 
Another feature of the region of a sudden transition from anti-correlation to correlation is that the semiring with zero values of $h_3$ coincides with the semiring of $h_4$ minima (\citealp{Zakharova_etal2023}, figure~A.3). Here, the 
disc and bar particles make an almost equal contribution to the total LOSVD, which turns out to be almost symmetric (zero $h_3$) but broadened (negative $h_4$).
It is true for all of our models (for the BL model, see Fig.~\ref{app:alternatingh3_LOSVDs_xbl}).
\par
Even closer to the centre \textcolor{black}{(the D stripe)}, the LOSVD is very wide, which is associated with a large velocity dispersion, the distribution is almost symmetric, with peaks near zero on the side of negative velocity values. The $h_3$ parameter has also negative values close to zero. A weak positive correlation $h_3-\overline V$ is observed here.
\par
We should note that the spots of correlations in the area where the bar adjoins the disc is also visible for the BL model, but at larger inclinations (Fig.~\ref{fig:app_bl_h3_full}, $i=60^\circ$, $\mathrm{PA}=0^\circ$; see also Fig.~\ref{fig:bl_and_blx_for_discussion}, right column). The nature of these spots is the same as for the X model. At tangent points, particles of the disc and bar are mixed. The bar particles create the main peak at low values of velocities, and the disc particles form a rapidly rotating wing. But for the BL model, only at $i=60^\circ$ does the line of sight capture enough bar particles in the ansae region to create a low-velocity peak and a correlation. Such a feature can be observed for barred galaxies with prominent ansae even at $i\approx40-60^\circ$.

\subsection{Barlens-specific signature nature}
\label{sec:bl_model_discussion}

\begin{figure}
   \centering
   \includegraphics[width=1\linewidth]{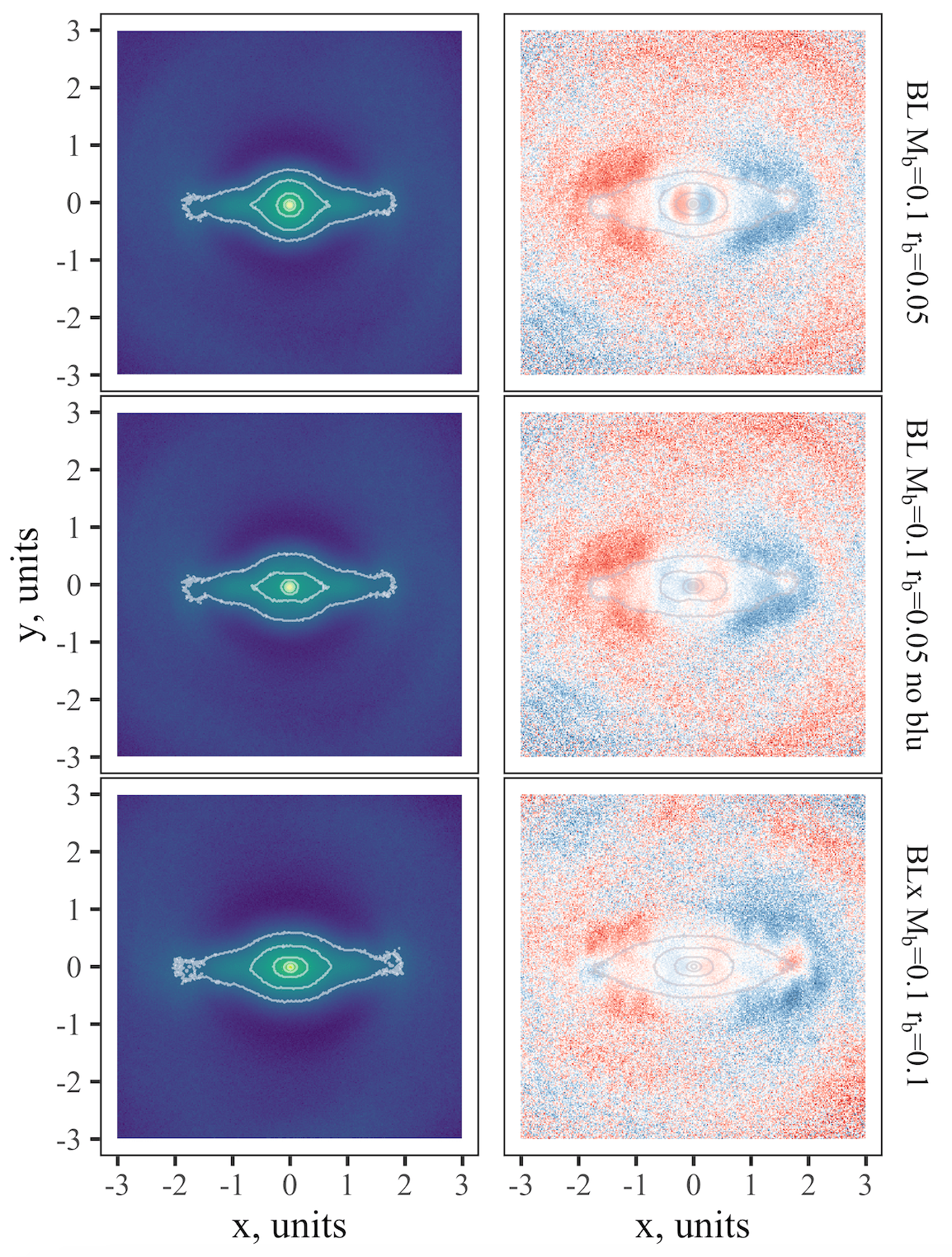}
   \caption{The comparison between two models with barlenses, the BL model with a compact bulge and the BLx model with a rarefied bulge. The characteristics of the bulge for each model are on the right side of the panel. The first row shows the intensity (\textit{left}) and $h_3$ maps (\textit{right}) for the full BL model. The second row represents the same for the BL model without \blu orbits. The third row shows the BLx model, which does not show the same as the BL model signature on the $h_3$ map. The intensity maps are shown for $i=0^\circ$, $\mathrm{PA}=0^\circ$ while the $h_3$ are for $i=40^\circ$, $\mathrm{PA}=0^\circ$. The panels in the first column contain isophotes, which correspond to the intensity maps at $i=0^\circ$ while isophotes of the second column correspond to the intensity at $i=40^\circ$.}
    \label{fig:bl_and_blx_for_discussion}%
\end{figure}

The central anti-correlation noted in Section~\ref{sec:barlens}, which is characteristic of our BL model, seems to be a feature of barlens models only with compact bulges. This feature is eliminated only by the exclusion of the \blu orbital family (Fig.~\ref{fig:bl_i40pa0_deleteorbits} and Fig.~\ref{fig:app_bl_no_blu_h3_full}), which is the main building block of the rounded barlens in the innermost parts of the model \citep{Smirnov_etal2021,Tikhonenko_etal2021}.
\par
Fig.~\ref{fig:bl_and_blx_for_discussion} (the first column) shows the face-on intensity maps for the BL, the BL model without \blu orbits, and the BLx models. The outer parts of the barlens stand out well in both models and assemble from orbits belonging to the \blo orbital family. This family is representative in both models (see Table~1 in \citealp{Smirnov_etal2021}). At the same time, the innermost parts of the barlens assemble from the \blu orbits, and the barlenses differ morphologically. In the BL model, this component stands out clearly by its rounded isophotes. The BLx model has a less rounded outline, and the \blu family is poorer represented here (3.4\% of all orbits in the BLx model versus 9\% in the BL model, Table~1 in \citealp{Smirnov_etal2021}). As a consequence, the BLx model does not show the central feature on the $h_3$ maps in the form of anti-correlations (Fig.~\ref{fig:bl_and_blx_for_discussion}, the third row, $i=40^\circ$, $\mathrm{PA}=0^\circ$). If \blu orbits are excluded from the BL model, then the $h_3$ maps  become very similar, featuring weak central correlations (Fig.~\ref{fig:bl_and_blx_for_discussion}, middle row).
\par
In \citet{Iannuzzi_Athanassoula2015}, there is a model gtr116, presented as a model with a barlens \citep{Athanassoula_etal2015}. However, on the $h_3$ map with $i=60^\circ$ and $\mathrm{PA}=90^\circ$ (figure~29 in \citealp{Iannuzzi_Athanassoula2015}), such a central feature is not visible, as well as on the $h_3$ map for our BLx model. It can be assumed that the gtr116 model does not have a sufficiently large number of orbits from the \blu orbital family~(i.e., gtr116 seems to be coupled with a more rarefied bulge and does not acquire this feature).
\par
There is one more, almost curious case when the $h_3-\overline V$ anti-correlation can be observed in the inner parts of a barred galaxy without a barlens. This feature is demonstrated by our X model at $i=60^\circ$ and $\mathrm{PA}=90^\circ$ (Fig.~\ref{fig:app_x_h3_full}, bottom right plot). There is no barlens, inner disc, or second inner bar \citep{Du_etal2016} in this model. This is a purely geometric effect. The angle of $60^\circ$ is a complementary angle to the opening angle of the B/PS bulge X structure \citep{Smirnov_Sotnikova2018}. In this case, at $\mathrm{PA}=90^\circ$, the line of sight slides along the ray of the B/PS bulge X structure, collecting many particles with a small projection of the rotation velocity. The situation is similar to that for the outer areas of the disc at the edge-on view. A wing to the right of the peak with a small negative velocity is added to the wide LOSVD for tangent points. Such a wing creates an anti-correlation of $h_3$ with velocity. 
In a certain range around $i=60^\circ$ the parameter $h_3$ becomes positive. At smaller and larger angles, when the line of sight slides off the ray of the X structure, it takes on negative values. One can compare plots in Fig.~\ref{fig:app_x_40_60} for $i=40^\circ$ and $i=60^\circ$. Orange points represent near tangent particles, and blue points are distant particles from higher altitudes. Blue particles for $i=60^\circ$ create a low-velocity wing that results in a weak anti-correlation\footnote{Such an effect is not observed for the BL model at $i=60^\circ$ and $\mathrm{PA}=0^\circ$ when the \blu orbits are removed. In this case, anti-correlation is not preserved, since in this case the geometric effect does not work (see Appendix~\ref{app:inclination_losvds} and Fig.~\ref{fig:blu_influence_on_LOSVDs_allmaps}).}. This is undoubtedly an interesting observation, but it is unlikely to have wide astrophysical applications.

\subsection{Application to real data}
\label{sec:obs_discussion}
\begin{figure*}
   \centering
   \includegraphics[width=1\linewidth]{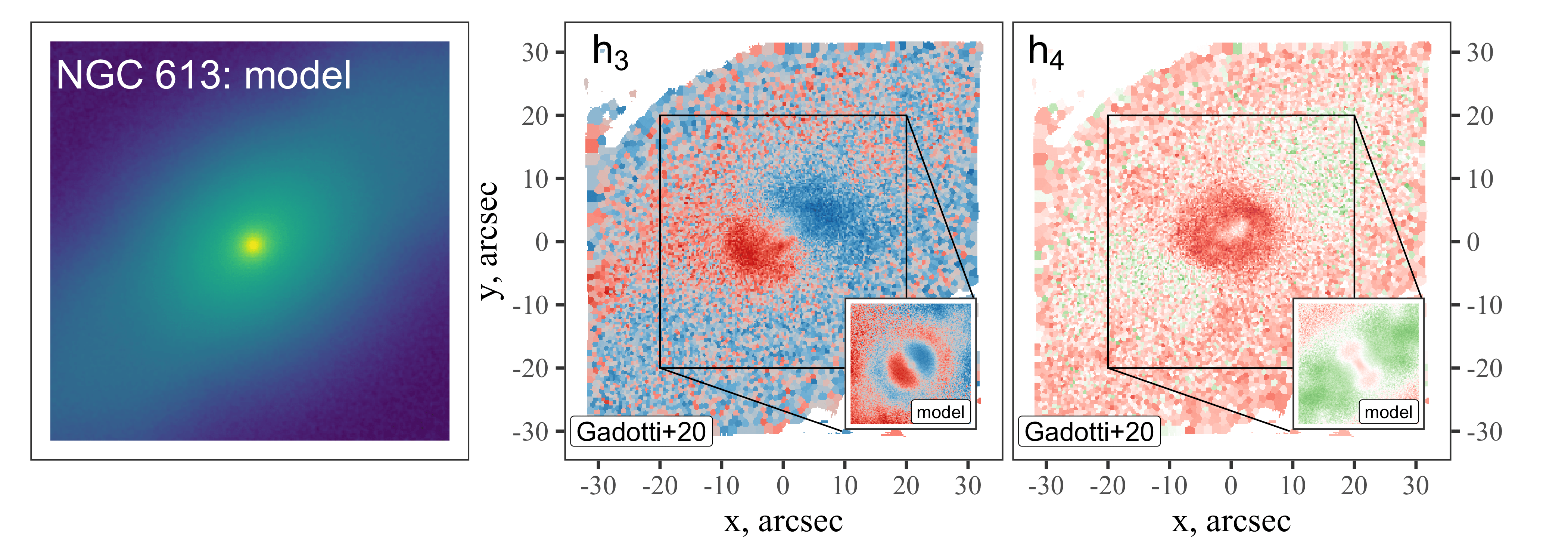}
   \includegraphics[width=1\linewidth]{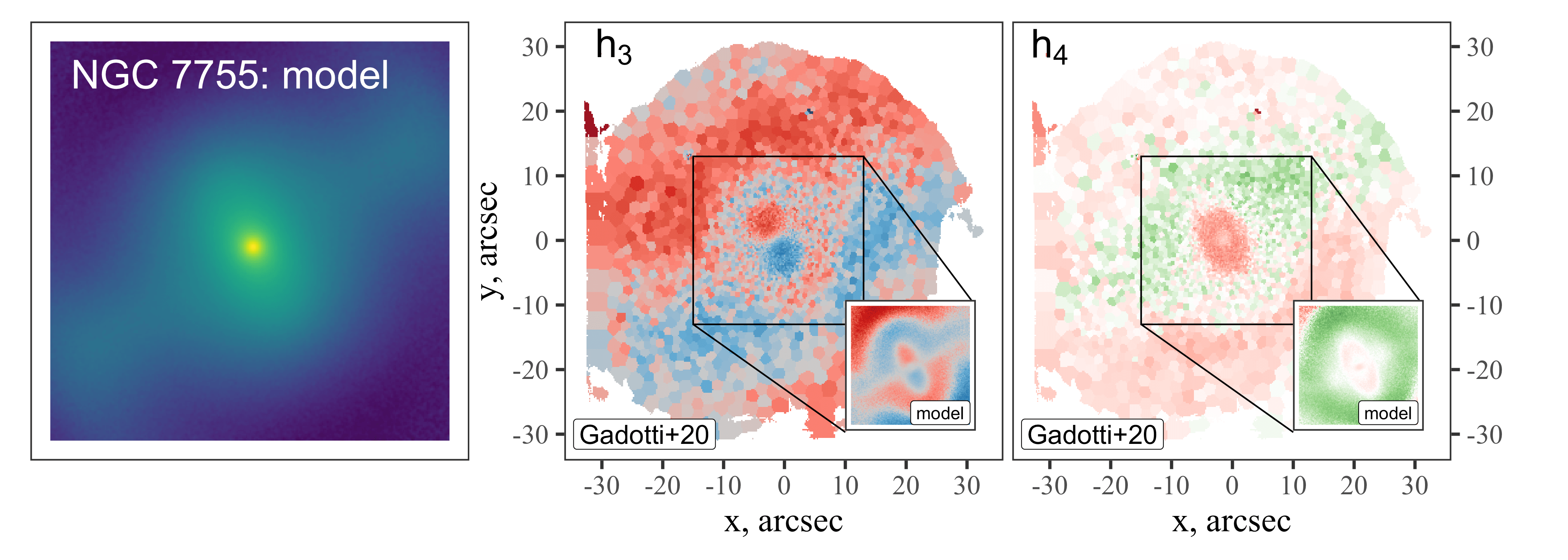}
   \caption{The BL model rotated to mimic two galaxies from the TIMER sample: NGC~613 (\textit{top row}, $i=39^\circ$, $\mathrm{PA}=1^\circ$) and NGC~7755 (\textit{bottom row}, $i=53^\circ$, $\text{PA}=-85^\circ$). \textcolor{black}{The bar viewing angles are borrowed from decompositions in the online supplement of \citet{Salo_etal2015} and corrected for inclination.}
   For each row, the \textit{left} panel shows the intensity map of the model. On the \textit{middle} panel we reproduce the map of $h_3$ parameter from \citet{Gadotti_etal2020} with a synthetic $h_3$ map of our BL model shown on the inset plot. On the \textit{right} panel there is the map of $h_4$ parameter from \citet{Gadotti_etal2020} with an inset of $h_4$ map based on our BL model. \textcolor{black}{The parameters $h_3$ and $h_4$ were calculated for observational data by \citet{Gadotti_etal2020}, and for our models by us, so we suggest paying attention only to the sign, and do not compare exact values.}}
    \label{fig:real_gals}%
\end{figure*}

\textcolor{black}{We now discuss the applicability of our models to observations. \citet{Gadotti_etal2020} provide kinematic maps ($\overline V$, $\sigma_\mathrm{LOS}$, $h_3$ and $h_4$) for a number of barred galaxies that were classified by \citet{Buta_etal2015} as galaxies with a barlens morphology (see also \citealp{Laurikainen_Salo2017}). Among the galaxies in \citet{Gadotti_etal2020}, we selected two galaxies for which our BL model is able to reproduce the features of the $h_3$ and $h_4$ maps simultaneously. These are the galaxies NGC~613 and NGC~7755.} 
\par
\textcolor{black}{\textit{NGC~613} is classified as a barlens galaxy \citep{Buta_etal2015,Laurikainen_Salo2017}. The galaxy has a nuclear disc and a star-forming nuclear ring \citep{Gadotti_etal2020}. S$^4$G photometric decomposition \citep{Salo_etal2015} does not include a barlens component and gives for the galaxy, in addition to the disc and bar, a bulge  with $B/T = 0.13$ and the Sersic index $n=0.80$, which is not typical for classical bulges. However, as shown by \citet{Laurikainen_etal2018}, with an accurate photometric multicomponent decomposition, the bulges of the initial rough decomposition (bulge+disc+bar) can be further split into a barlens as a separate photometric component and a central compact bulge.
Therefore, we assume that NGC~613 may be similar to our BL model, which means it should have similar $h_3/h_4$ maps with the BL model. NGC~613 represents a specific case of$\mathrm{PA}\approx 0^\circ$. We rotated our BL model according to the viewing angles of this galaxy \citep{Salo_etal2015}, 
and reconstructed kinematic maps. Fig.\ref{fig:real_gals} (top row) shows the observational data from \citet{Gadotti_etal2020} and our simulated data. The first column (plot) in the figure shows snapshots of the BL model for corresponding viewing angles, the second and the third ones show $h_3/h_4$ maps.}
\par
\textcolor{black}{On the $h_3/h_4$ maps, NGC~613 demonstrates characteristic features for $\mathrm{PA}=0^\circ$. One can see the central $h_3 - \overline V$ anti-correlation along the major axis of the galaxy and areas of negative $h_4$ values outside the barlens on the major axis. We associate these minima in the region, where the narrow bar begins, with the orbits of the x1 family, protruding in the vertical direction and making the main contribution to the B/P shape \citep{Zakharova_etal2023}}
\par
\textcolor{black}{\textit{NGC~7755} exhibits a clear barlens morphology in the S$^4$G images, although it is not included in the catalog by \citet{Buta_etal2015}. The galaxy has also a nuclear disc and a ring \citep{Gadotti_etal2020}. The ring of H$\alpha$ emission at its outermost edge is embedded in a region of older stellar populations \citep{Bittner_etal2020}. S$^4$G photometric decomposition \citep{Salo_etal2015} includes only a disc, bar and bulge with $B/T = 0.16$ and a Sersic index $n=0.65$, which is not typical for classical bulges. We also assume that our BL model is suitable for describing this galaxy.}
\par
\textcolor{black}{NGC~7755 represents another specific case in terms of viewing angle, with $\mathrm{PA}\approx -90^\circ$. Fig.\ref{fig:real_gals} (bottom row) shows the observational data from \citet{Gadotti_etal2020} and our simulated data (inset pictures) for this galaxy. Now one can see the central $h_3 - \overline V$ anti-correlation along the minor axis of the galaxy (middle plot) and the ring of negative $h_4$ values (right plot) inside which the barlens is located. We note that the deep $h_4$ minima inside this ring occur at almost zero values of $h_3$ in the transition region from the B/PS bulge to the disc and delineate precisely the boundary of the B/PS, and not the barlens itself. These minima are associated not with the peculiarities of the vertical density distribution (peanuts) but with the contamination of the LOSVD with various velocity components and the mixing of LOSVDs from different orbital groups \citep{Zakharova_etal2023}.}
\par
Thus, our BL model is able to reproduce the observed data, especially for the $h_3$ parameter, and must also be taken into account as a possible explanation of the observed features.
\subsection{Caveats: nuclear disc/inner bars vs barlens}
\label{sec:inner_disc_barlens}
We note the limitations of our models. Our models do not include gas particles and, accordingly, can not reproduce, for example, the formation of any disc-like component due to star formation in the central region.  
As a consequence, we can calculate the velocity dispersion only for those components that were originally specified or formed as a result of the bar instability.
Therefore, we rely only on the $h_3/h_4$ parameters, avoiding $v/\sigma$ ratios commonly used in the literature (e.g., \citealp{Gadotti_etal2020}). However, as shown above, our models reproduce both the features of models with gas from other works (e.g., \citealp{Iannuzzi_Athanassoula2015}) and observational data, although they are less demanding on computing power. We also point out that the central anti-correlation of the BL model explains the signature on the corresponding maps of real galaxies with possible barlenses. In previous work~\citep{Zakharova_etal2023}, we tried to distinguish the barlens on $h_4$ maps and have come to the conclusion that $h_4$ maps only show the signature of the B/PS bulge and do not reveal the exact face-on bar morphology. In order to identify the barlens model in our grid of models, it is precisely the $h_3$ parameter that is relevant. 
\textcolor{black}{We have shown that the model with a barlens has a $h_3$-$\overline{V}$ anticorrelation in the central part and barlens-typical orbits responsible for this signature.} \textcolor{black}{Although the barlens model is capable of explaining central anti-correlation, it is not the only solution. Central circular orbits can also be associated with inner bars or nuclear discs.}
\par
\textcolor{black}{Inner bars demonstrate a strong $h_3-\overline V$ anti-correlation in central regions. At the same time, they exhibit a characteristic feature on $\sigma_\mathrm{LOS}$ maps, namely $\sigma$-humps along the minor axis of an inner bar and $\sigma$-hollows along its major axis \citep{deLorenzo_Caceres_etal2008,Du_etal2016}. For our BL model we found only} minima in velocity dispersion radial profile at the very centre of the model (similar to \citealt{Bureau_Athanassoula2005}). However, these minima occur on much smaller scales ($r \approx 0.05$) than the size of the barlens ($0.5$). \textcolor{black}{Moreover, we did not introduce any additional features into our model, such as a rotating bulge \citep{Debattista_Shen2007}, a dynamically cold central disc \citep{Du_etal2015}, or a very massive dark halo \citep{Saha_Maciejewski2013}, which would produce inner bars in models without gas. As a consequence, there is no inner bar in our BL model.}
\par
\textcolor{black}{\citet{Tikhonenko_etal2021} showed that the structure assembled from \blu orbits resembles a rather flat component surrounded by a `halo' (see their figure~5, bottom plot). This flat component can look like a nuclear disc and manifest itself kinematically as a central disc. The inner flat part of a barlens is an extended structure and its half-mass radius is one fourth of the radial scale of the disc \citep{Tikhonenko_etal2021}. 
Although nuclear discs can also have sizes up to 1~kpc \citep{Gadotti_etal2020}, the barlens in our model is formed from the same material as a bar, so the composition of their stellar populations is similar. This may distinguish barlenses from nuclear discs which are generally composed from a younger stellar population than their surroundings \citep{Bittner_etal2020}.
In any case, we note that $N$-body/hydrodynamics simulations are needed to make more subtle distinctions between the central part of the barlens and the nuclear disc.}

\section{Conclusions}
\label{sec:conclusions}
In two works, current and \citet{Zakharova_etal2023}, we have analysed our galaxy models with B/PS bulges, which reproduce the observational distortion of the LOSVD characteristics of the B/PS bulges, and for the barlens especially. We have shown that the galaxies with a barlens~(formed due to the presence of a compact bulge) can be described by central $h_3-\overline V$ anti-correlation and negative ring of $h_4$ parameter that surrounds the anti-correlation region\footnote{Or at least the anti-correlation region is accompanied by minima of $h_4$ on the major axis \citep{Zakharova_etal2023}.}. Moreover, the anti-correlation is a specific signature of the barlens~(and exact barlens-specific orbits responsible for it), but parameter $h_4$ reflects mostly the peculiarities on the LOSVD, which is characteristic not only for the barlens but for all our B/PS bulges. 
Therefore, in the case of ambiguity in the signatures of the $h_3$ and $h_4$ parameters,  \textcolor{black}{it is better to use direct LOSVD analysis for individual pixels instead of high-order LOSVD moments.} This applies to those cases when \textcolor{black}{unambiguous signs of kinematically different} subsystems are visible in the LOSVD.
\par
The other results of this work are as follows:
\begin{itemize}
    \item the barlens coupled with a rarefied bulge~
    (BLx model) does not show a central $h_3-\overline V$ anti-correlation on the $h_3$ maps;
    \item the morphology of our models has a clear imprint on the $h_3$ maps and can help in bar identification;
    \item such identification is especially useful at $\mathrm{PA}=90^\circ$ and large $i$: a change of anti-correlation to correlation while moving towards the centre along the major axis tells about the presence of BP/S bulge and sets its boundary;
    \item the mechanisms of $h_3$ maps formation either relate to the type of orbits~(elongated or circular) or to the mixing of components with different typical velocities~(as, for instance, in the areas where the bar connects to the disc).
\end{itemize}
\par

\section{Acknowledgements}
\textcolor{black}{We thank the anonymous referee for the comments, that helped to improve the quality of the presentation of our results. }   
The authors express gratitude for the financial support from the Russian Science Foundation (grant no. 22-22-00376).
We are also grateful to the TIMER team \citep{Gadotti_etal2019} for making their data publicly available. This work made use of Astropy\footnote{\url{http://www.astropy.org}}: a community-developed core Python package and an ecosystem of tools and resources for astronomy \citep{astropy:2013, astropy:2018, astropy:2022}, as well as NEMO stellar dynamics toolbox \citep{Teuben_1995}. The authors also thank Roberto Saglia for a careful read of  a preliminary version of this manuscript and several suggestions on improving the text of the article. 

\section{Data availability}
The kinematic maps from the TIMER project are publicly available on the TIMER website. 
The rest of the data underlying this article can be shared on a reasonable request to the corresponding author.

\bibliographystyle{mnras}
\bibliography{bibliography} 

\appendix

\section{Entire $h_3$ maps for X and BL models}
\label{app:h3maps_full_inclinations}

\begin{figure}
   \centering
   \includegraphics[width=1\linewidth]{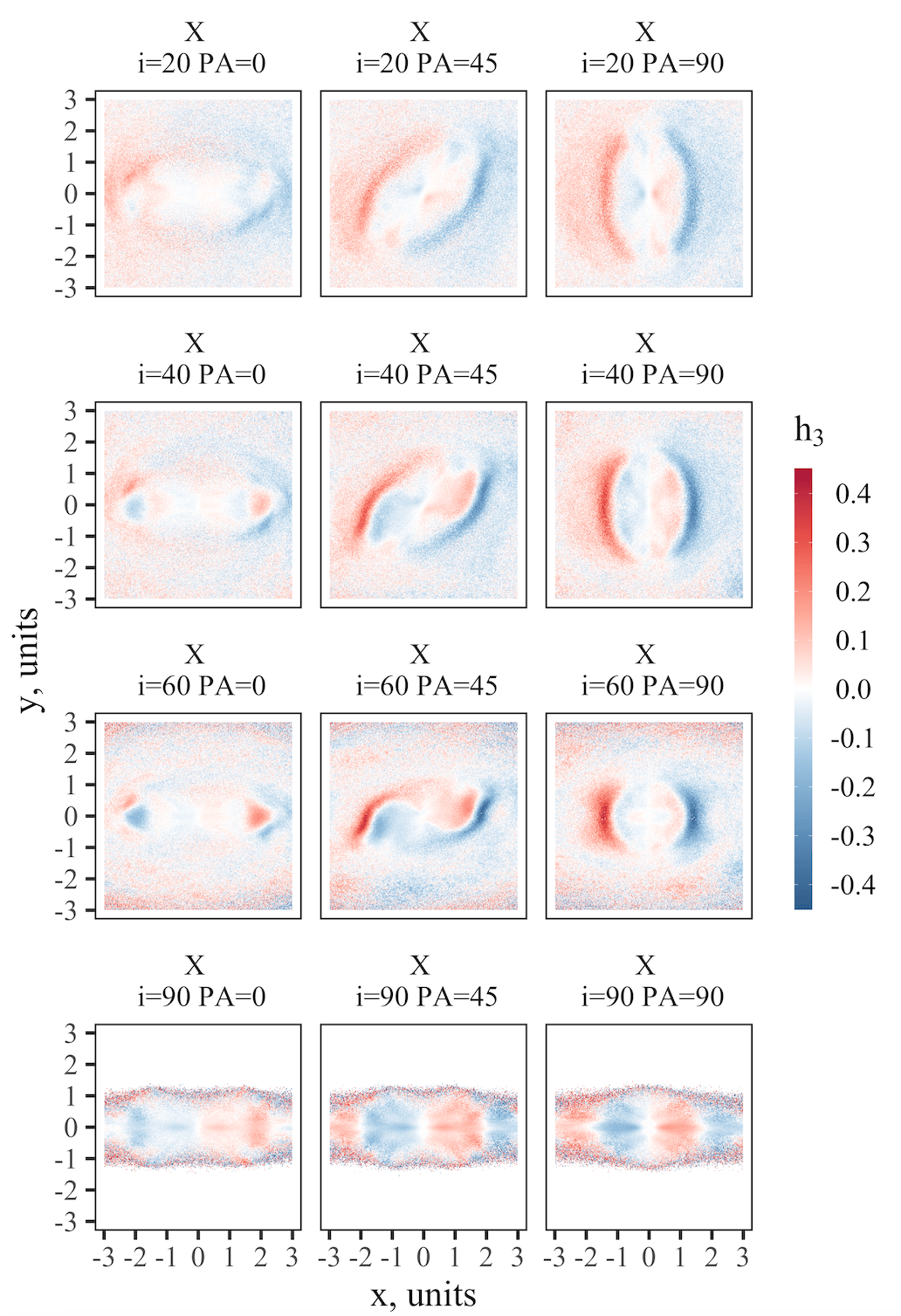}
   \caption{The $h_3$ maps for the X model with a wide range of inclinations and bar viewing angles.}
    \label{fig:app_x_h3_full}%
\end{figure}

\begin{figure}
   \centering
   \includegraphics[width=1\linewidth]{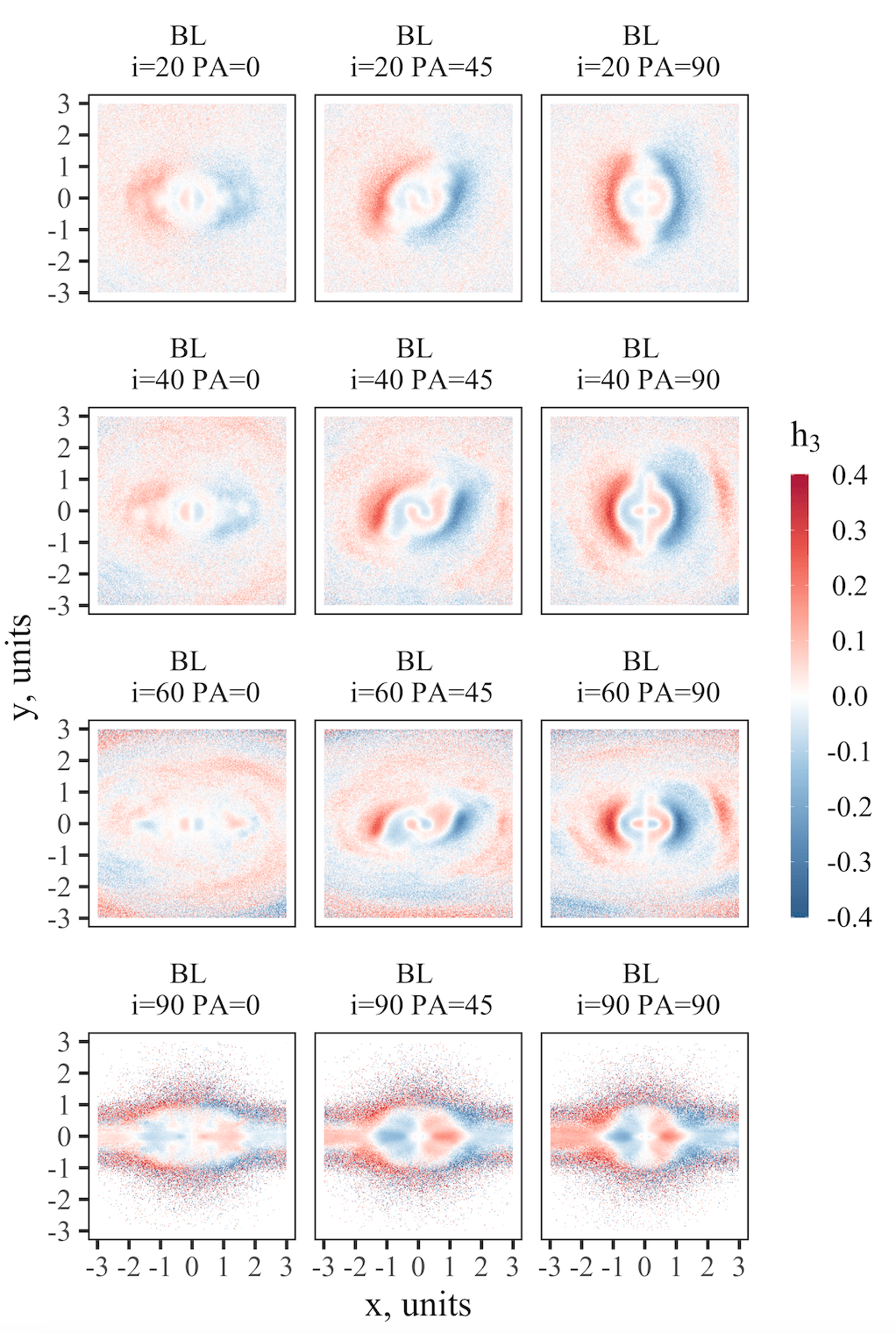}
   \caption{The same as in Fig.~\ref{fig:app_x_h3_full} but for the BL model.}
    \label{fig:app_bl_h3_full}%
\end{figure}
In Section~\ref{sec:inclination}, primarily $i=40^\circ$ case for the X and BL were discussed. In this section, we provide a full grid of all studied galaxy viewing angles for both models in Fig.~\ref{fig:app_x_h3_full} and Fig.~\ref{fig:app_bl_h3_full}, respectively. The inclination of models makes the $h_3$ features more prominent~($|h_3|$ increases) with maximal absolute $h_3$ values at $i \approx 60^\circ$. 
\par
Regardless of the orientation of the model, the $h_3$ maps follow the size of the bar~(the X model has a bigger bar than the BL model, and it can be traced in the figures).
\par
The connection of the bar to the disc in the ansae area noted by us is visible at all inclinations at $\mathrm{PA}=0^\circ$ for the X model. For the BL model, this feature is hardly distinguishable at $i<60^\circ$, but it is also always present.

\section{BL maps with excluded \blu orbits}
\label{app:h3maps_bl_noblu_inclinations}

\begin{figure}
   \centering
   \includegraphics[width=1\linewidth]{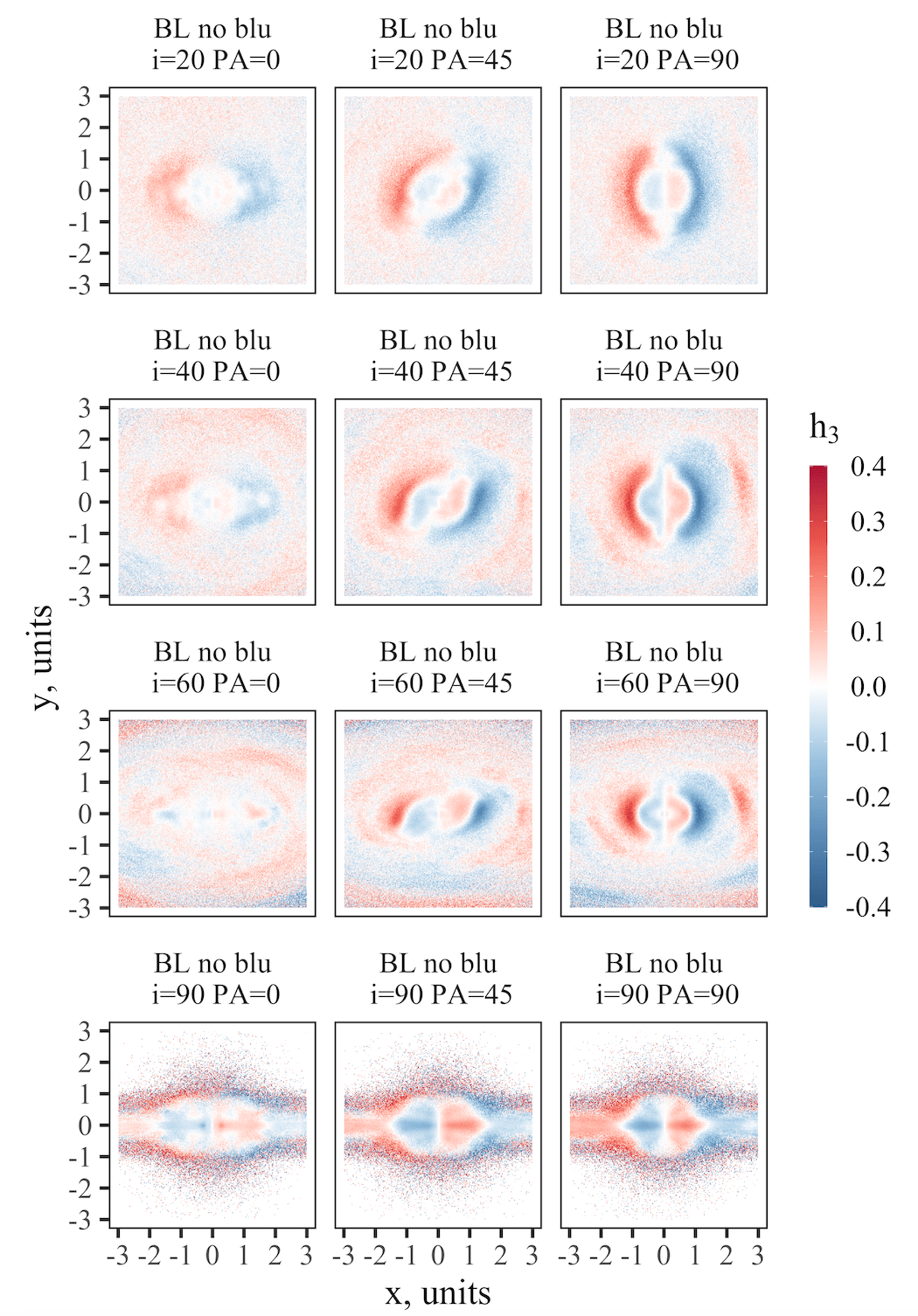}
   \caption{The $h_3$ maps for the BL model when \blu orbital group is excluded for all inclinations and PAs.}
    \label{fig:app_bl_no_blu_h3_full}%
\end{figure}

We establish that \blu orbits~(barlens-specific orbits) are responsible for the central $h_3-\overline V$ anti-correlation for the BL model using the removal of the signal of these orbits from the cube data. We now provide a full grid of inclinations and bar viewing angles to support our conclusion that the removal of these orbits removes the anti-correlation at any inclination and bar viewing angle. Fig.~\ref{fig:app_bl_no_blu_h3_full} shows $h_3$ maps for the BL model when \blu orbits are excluded. Each of these maps does not show the central $h_3-\overline V$ anti-correlation since we excluded the main barlens orbits.

\section{Position-velocity diagrams}
\label{app:losvds_major_axis}
Here, we provide the extended set of plots that demonstrate the features of LOSVDs along the LON for both X and BL models as in Fig.~\ref{fig:x_velocityanalysis_for_discussion}. The plots simultaneously demonstrate the positions of the LOSVDs peaks and the underlying different subsystem~(barlens/bar/disc) particles that contribute to the LOSVDs in each pixel. We also provide the $h_3$ values with stripes on the background layer as a reference. At $\mathrm{PA}=90^\circ$, one can see the abrupt transition from the B/PS bulge to the disc where $h_3-\overline V$ correlation changes to the anti-correlation.
\begin{figure*}
   \centering
   \includegraphics{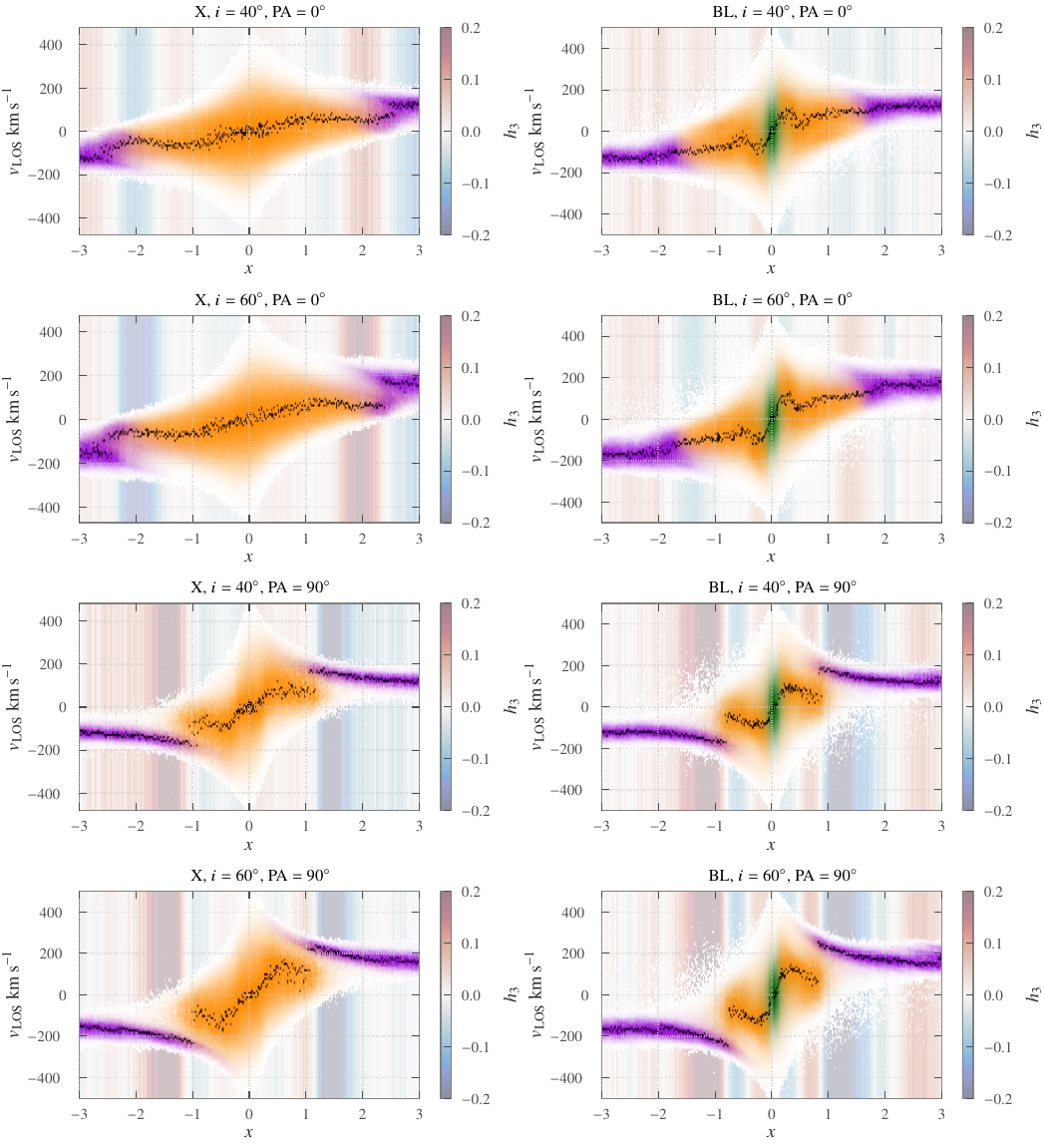}
 \caption{The features of LOSVDs for the X (left column) and BL (right column) models viewed from $i=40^\circ\, , 60^\circ$ and $\mathrm{PA}=0^\circ$ (two upper plots) and $\mathrm{PA}=90^\circ$ (two bottom plots). All symbols and colors are the same as in Fig.~\ref{fig:x_velocityanalysis_for_discussion}. The green dots on the right plots correspond to the LOSVDs of the bulge particles.}
 \label{app:alternatingh3_LOSVDs_xbl}
 \end{figure*}

\section{Effects of inclination on the $h_3$ - $\overline{V}$ relation in the inner parts of models}
\label{app:inclination_losvds}
Here, we explain how the LOSVD in a given pixel is formed at $\mathrm{PA}=90^\circ$ using both of our models as examples.
\par
\begin{figure*}
   \centering
   \includegraphics{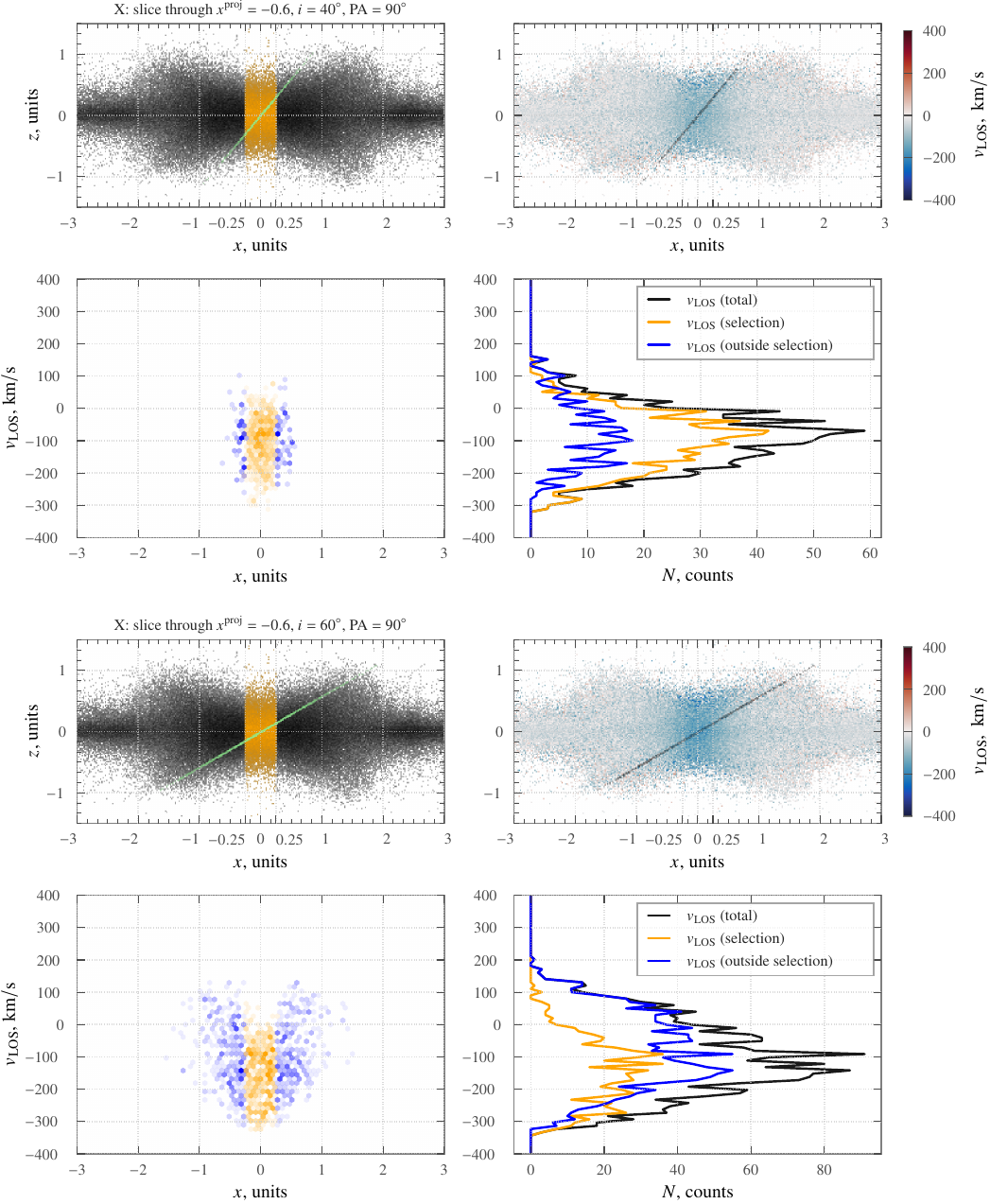}
   \caption{The formation of the LOSVD in a certain pixel at LON for $\mathrm{PA}=90^\circ$ (the X model). Two sets of four plots for $i=40^\circ$ and $i=60^\circ$ are shown. In each set, \textit{top left}: an intensity map of $xz$ slice ($x^{\mathrm{proj}} = y =-0.6$) containing the LOS (indicated with a green line); particles within $|x|<0.2$ are highlighted with orange color; \textit{top right}: the map of LOS velocity for the same slice with the LOS shown as a gray line; \textit{bottom left}: 2d LOSVD against $v_\mathrm{LOS}$ and $x$ coordinate; \textit{bottom right}: LOSVDs for all particles along the LOS (black), for $|x|<0.2$ particles only (orange), and for the rest of particles (blue), effectively, a marginal density for the one to the left.}
    \label{fig:app_x_40_60}%
\end{figure*}

For the X model, Fig.~\ref{fig:app_x_40_60} provides two sets of plots of four panels in each set for $\mathrm{PA} = 90^\circ$, $i=40^\circ$ and $i=60^\circ$. The upper left panel in each set shows the intensity for a thin slice selected by $x^{\mathrm{proj}} = -0.6$, which for the given viewing angle corresponds to a slice in the $xz$ plane near $y=0.6$. Orange color refers to the particles within $|x|<0.25$ (core particles in the notation by \citealp{Li_etal2018}). The green dotted line is the line of sight corresponding to the inclination $i$. The upper right plot demonstrates the values of $v_\mathrm{LOS}$ for a given inclination throughout the slice with the selected LOS indicated by the black dotted line.
The bottom row of each panel group shows the LOSVD, corresponding to the given inclination in a specified slice. The left plot is a two-dimensional distribution against both $x$ and $v_{\mathrm LOS}$, while the right one is a one-dimensional marginal density form of it, depending on the LOSVD only. The orange color is used to indicate the LOSVDs of the particles inside $|x|<0.25$ lane, and the blue color is assigned to the rest of the particles, most of which are located higher from the disc plane. 
The sum of two LOSVDs results in a different total LOSVD for $i=40^\circ$, and $i=60^\circ$.
The primary difference between these two inclinations is that the mean location of the orange ``band''  is shifted to higher velocities. It happens due to the increasing contribution of the azimuthal velocity component (which is dominant for these particles) to the LOSVD with increasing inclination. At the same time, the blue part of the LOSVD becomes more populated because more particles from the further parts of the galaxy fall off the LOS for $i=60^\circ$. 
Thus, for $i=40^\circ$, the main peak is associated with orange particles. The LOSVD for them is elevated from the side of high velocities, which gives a negative value of $h_3$. Since the value of $\overline V$ is also negative, this means a correlation. 
For $i=60^\circ$, the main peak falls on the blue particles from the ray of the B/PS bulge X structure. These particles populate the low-velocity wing of the LOSVD, resulting in anti-correlation.
\par
\begin{figure*}
   \centering
   \includegraphics{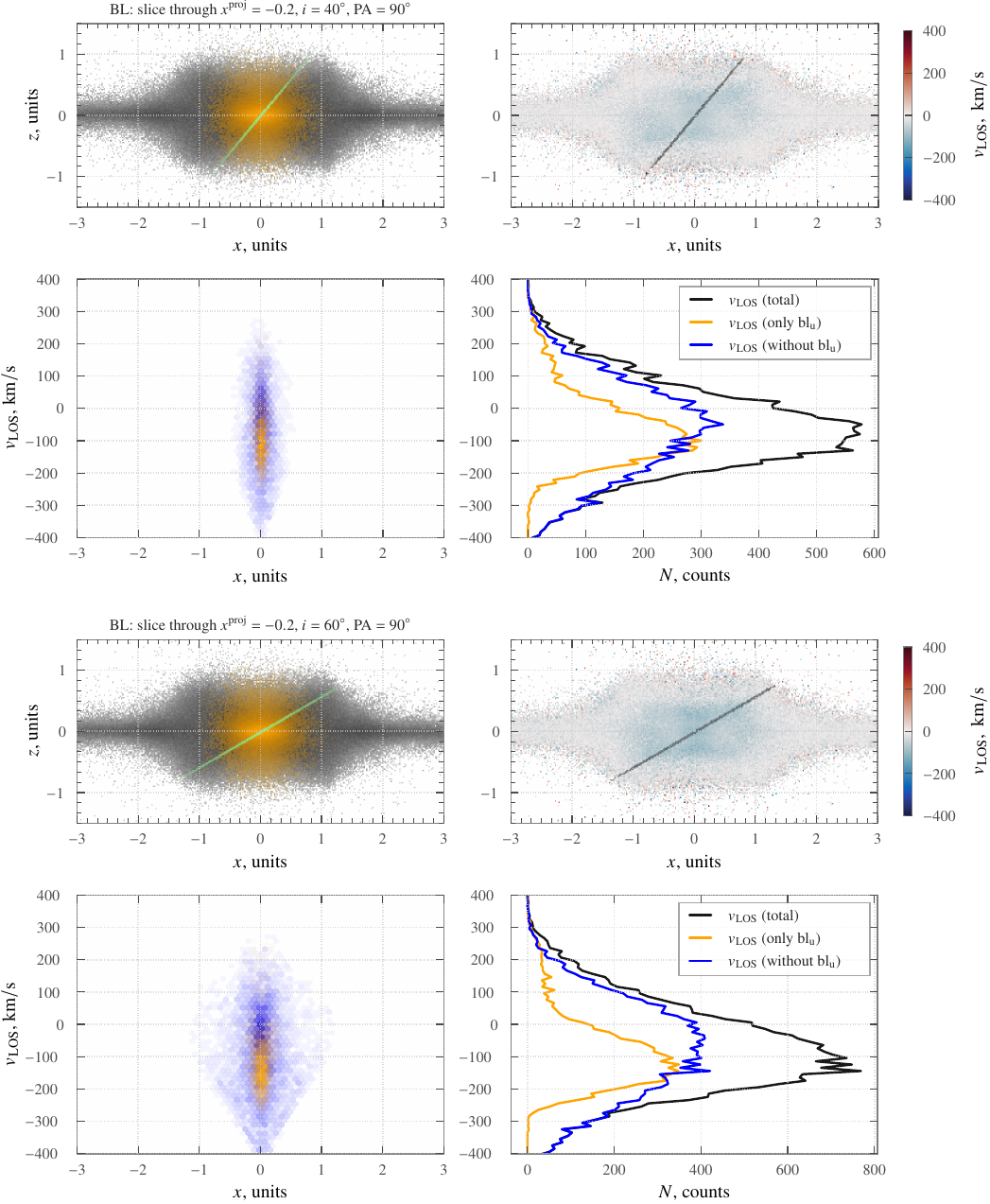}
   \caption{The same as in Fig.~\ref{fig:app_x_40_60} but for the BL model. The slice was taken at $y=-0.2$. Particles from the \blu orbital family are highlighted with orange color.}
   \label{fig:blu_influence_on_LOSVDs_allmaps}%
\end{figure*}

The notations in Fig.~\ref{fig:blu_influence_on_LOSVDs_allmaps}, which refers to the BL model, are the same as in Fig.~\ref{fig:app_x_40_60} except that the orange dots represent particles from the \blu orbital family. Here, we can see a completely different morphology of the B/PS bulge in a section in the $xz$ plane. It is not peanut-shaped but rectangular. At both $i=40^\circ$ and $i=60^\circ$, the line of sight picks up quite a lot of blue particles. The LOSVD associated with these particles has a low-velocity peak and is very wide. However, the shape of the total LOSVD is 
formed by orange \blu particles, which move rather fast (have negative values of velocity).
This results in a positive value of $h_3$ (anti-correlation).
At the same time, the LOSVD associated only with blue particles has a slightly elevated wing on the side of particles with higher velocities than peak velocities. If the orange particles (\blu orbits) are removed, this will lead to small negative values of $h_3$, i.e., to correlation (see Figs.~\ref{fig:bl_i40pa0_deleteorbits}, \,
\ref{fig:bl_i40pa0_deleteorbits_nod}, \,
\ref{fig:bl_and_blx_for_discussion} and \ref{fig:app_bl_no_blu_h3_full}). This works the same way for $i=40^\circ$ as it does for $i=60^\circ$. Therefore, at $i=60^\circ$, there is no curious effect of anti-correlation as for the X model when the \blu orbits are removed.

\bsp	
\label{lastpage}
\end{document}